\documentclass[prx,showpacs,twocolumn,amsmath,amssymb,floatfix]{revtex4-1}

\usepackage{hyperref}
\usepackage{graphicx}
\usepackage{bm}
\usepackage{color}
\usepackage[applemac]{inputenc}
\usepackage{amssymb}

\begin{document}  
   
\title{Measuring Majorana non-locality and spin structure with a quantum dot}

\author{Elsa Prada$^1$, Ramón Aguado$^2$, Pablo San-Jose$^2$}
\affiliation{$^1$Departamento de Física de la Materia Condensada, Condensed Matter Physics Center (IFIMAC) and Instituto Nicolás Cabrera, Universidad Autónoma de Madrid, E-28049 Madrid, Spain\\
$^3$Instituto de Ciencia de Materiales de Madrid, Consejo Superior de Investigaciones Científicas (ICMM-CSIC), Sor Juana Inés de la Cruz 3, 28049 Madrid, Spain}

\date{\today} 

\begin{abstract}
Robust zero bias transport anomalies in semiconducting nanowires with proximity-induced superconductivity have been convincingly demonstrated in various experiments. While these are compatible with the existence of Majorana zero modes at the ends of the nanowire, a direct proof of their non-locality and topological protection is now needed. Here we show that a quantum dot at the end of the nanowire may be used as a powerful spectroscopic tool to quantify the degree of Majorana non-locality through a local transport measurement. Moreover, the spin polarization of dot sub-gap states at singlet-doublet transitions in the Coulomb blockade regime allows the dot to directly probe the spin structure of the Majorana wave function, and indirectly measure the spin-orbit coupling of the nanowire.
\end{abstract}

\maketitle

\section{Introduction}
Majorana zero modes, peculiar self-conjugate Bogoliubov quasiparticles that emerge at zero-energy in topological superconductors, may one day constitute the building blocks of topologically-protected quantum computation 
\cite{Nayak:RMP08,Sarma:NQI15}. Arguably, the simplest way to obtain such exotic quasiparticles is to artificially engineer topological superconductivity by means of the superconducting proximity effect \cite{Fu:PRL08}. One of the most promising routes using this idea is based on proximitized semiconducting nanowires with strong spin-orbit coupling and in the presence of an external magnetic field \cite{Lutchyn:PRL10,Oreg:PRL10}. Indeed, several experiments have reported zero bias transport anomalies in such proximitized nanowires \cite{Mourik:S12,Deng:NL12,Das:NP12,Churchill:PRB13,Lee:NN14,Zhang:16,Deng:S16}. The observed anomalies are in some cases remarkably robust \cite{Zhang:16,Deng:S16}, as expected of zero modes of topological origin, and are interpreted as evidence of Majorana zero modes and induced topological superconductivity in the nanowires. While these results are highly promising, it is now necessary to obtain direct evidence of the crucial property of Majoranas that underlies their protection: spatial non-locality. 

\begin{figure}[t!]
   \centering
   \includegraphics[width=\columnwidth]{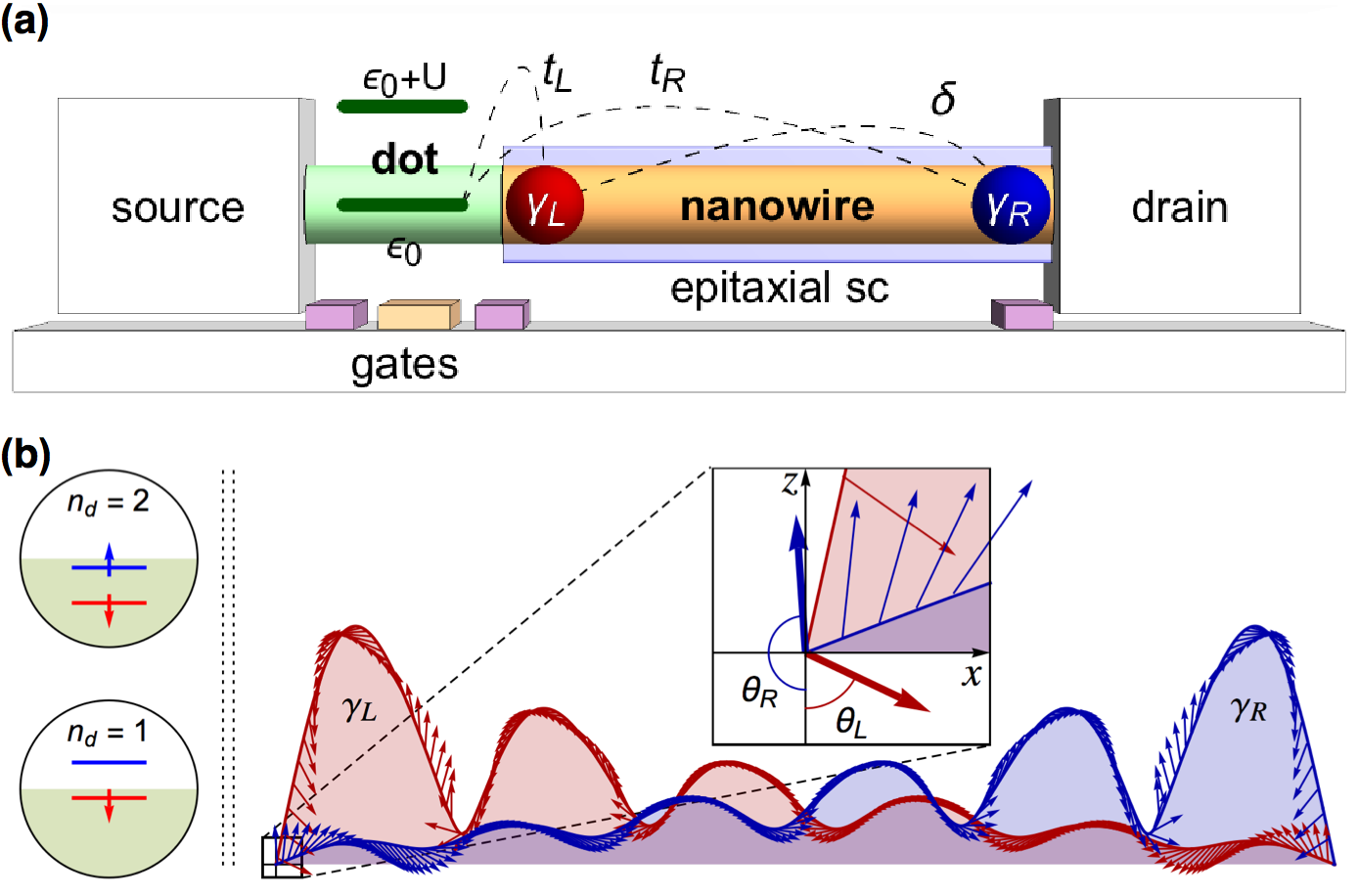}
   \caption{(a) Sketch of a quantum dot-nanowire junction proposed to measure the Majorana's degree of non-locality and spin structure. Tunnelling differential conductance across the dot provides access to the coupled dot-nanowire spectrum. As the quantum dot levels (spin polarized due to Coulomb blockade) are tuned across zero energy with a local gate, they anticross the Majorana zero modes $\gamma_L$ and $\gamma_R$, with a strength that depends on the relative spin orientation of the Majorana and the dot states. [The spin texture of the two Majoranas and the relevant spin canting angles $\theta_{L,R}$ are shown in panel (b)]. The anticrossings thus provides direct access to the Majorana spin structure at the contact. In contrast, the energy of the two Majorana bound states stays pinned at zero at the crossings if they do not overlap, i.e. if their hybridization $\delta$ is zero and the dot is only coupled to $\gamma_L$ (hopping $t_L$ is finite but $t_R$ is zero). Deviations from such non-locality condition manifest as particular $\epsilon_0$-dependent energies of the hybridized Majoranas, see Fig. \ref{fig:levels}, that may be used to quantify the amount of Majorana overlap as $\Omega \approx \sqrt{t_R/t_L}$. }
   \label{fig:sketch}
\end{figure}

A direct demonstration of Majorana non-locality would immediately rule out non-topological origins of the observed zero bias anomalies, such as disorder \cite{Liu:PRL12,Pikulin:NJP12}, or various effects related to unintentional quantum dot formation in the nanowire such as the Kondo effect \cite{Lee:PRL12,Zitko:PRB15} or Zeeman-induced sub-gap states \cite{Lee:NN14}. The challenge seems daunting however. A great variety of sophisticated schemes have been devised for probing non-locality of Majoranas and their non-Abelian braiding properties \cite{Nilsson:PRL08,Tewari:PRL08,Fu:PRL10,Alicea:NP11,Sau:PRB11,Halperin:PRB12,Heck:NJP12,Hyart:PRB13,Liu:PRB13,Zocher:PRL13,Fregoso:PRB13,Li:SR14,Lobos:NJP15,Haim:PRL15,Aasen:PRX16,Chiu:EEL15,Vijay:PRB16,Karzig:16,Plugge:NJP17}.
Here we show that the problem of quantifying Majorana non-locality might have a simpler solution than anticipated, requiring only a local probe. By measuring transport into the Majorana nanowire across an intervening quantum dot, as those that often form when gating the nanowires (Fig. \ref{fig:sketch}), Majorana non-locality may be unambiguously demonstrated. This result suggests that quantum dots are unexpectedly powerful spectroscopic tools to fully characterize essential Majorana properties. These extend beyond spatial non-locality and include the spin structure of the Majorana wave function at the end of the nanowire, that can be accessed rather simply by virtue of the spin-polarization of the dot states due to Coulomb blockade. Measuring the Majorana spin structure using such a spin-selective probe also allows to indirectly extract the spin-orbit coupling in the nanowire, providing a complementary measurement to more conventional techniques \cite{Mishmash:PRB16,Deng:S16}.

The structure of the paper is the following. We first study a microscopic tight-biding model, Sec. \ref{sec:model}, for a dot in the Coulomb blockade regime coupled to a finite-length topological superconductor nanowire, and characterize numerically its spectral phenomenology in Sec. \ref{sec:TB}. To connect the numerical results of the anticrossings to the physical quantities of interest, we develop a simple low-energy effective model in Sec. \ref{sec:effmodel} with which we interpret the dot-Majorana anticrossings in Sec. \ref{sec:int}. In Sec. \ref{sec:nonlocal} we derive an estimator that can quantify the degree of Majorana non-locality by local measurements, and analyse its behaviour under generalizations of the nanowire model in Sec. \ref{sec:beyond}. Finally, we derive analytical formulae relating the microscopic parameters to the effective model parameters in Appendix \ref{sec:micro}. The main results of our analysis, discussed in the concluding Sec. \ref{sec:conclusions}, can be condensed as follows:

\begin{itemize}
\item Majorana non-locality, and hence topological protection, is revealed as a zero-energy mode that does not shift as the two dot levels cross zero energy (Figs. \ref{fig:TBlong}d and \ref{fig:levels}a). 
\item In contrast, the dot states avoid the zero mode at resonance. A comparison of the anticrossing strength of the two spin-polarized dot levels directly yields the degree of spin canting of the Majoranas at the end of the nanowire.
\item In the presence of a finite overlap between the Majoranas, their energy shifts away from zero and follows a bowtie or diamond-like pattern around the dot level resonances (e.g. Figs. \ref{fig:TBshort}b,c and \ref{fig:levels}b,c). The details of this pattern can be used to obtain an accurate estimator $\Omega\approx\sqrt{t_R/t_L}$ of the degree of non-locality of the Majoranas and of their expected immunity against decoherence from local noise.
\item The dependence of the spin canting angle with Zeeman field can be used to indirectly measure the spin-orbit coupling in the nanowire.
\end{itemize}

\section{Tight-binding dot-nanowire model}
\label{sec:model}

An interacting quantum dot coupled to a superconducting contact is an artificial analogue of a quantum impurity in a superconductor. The physics of such hybrid device is governed by the fermionic parity and spin of the two possible ground states, doublet or singlet, and their corresponding sub-gap excitations (which are sometimes called Shiba states when they are spin-polarized). Here we study in detail a generalisation of this paradigmatic model where the superconductor is replaced by a proximitized semiconducting nanowire which becomes a topological superconductor for large enough Zeeman fields \cite{Lutchyn:PRL10,Oreg:PRL10}. Such a quantum dot-topological superconductor junction can be experimentally realized by creating quantum dots at the end of the nanowire using e.g. depleting gates. 
In this section we study numerically, using a microscopic tight-binding model, the hybridisation between the Shiba subgap states in the quantum dot and the Majorana zero modes that appear in the nanowire in the topological phase. 

Consider a quantum dot with a single spinful level coupled to the left end ($x=0$) of a proximitized Rashba nanowire of length $L_w$ under a Zeeman field $B$, see sketch in Fig. \ref{fig:sketch}. A rather general model for the system (see Sec. \ref{sec:beyond} for extensions) reads \cite{Lutchyn:PRL10,Oreg:PRL10}
\begin{eqnarray}
H&=&H_d+H_w+H_\mathrm{hop} \nonumber\\
H_d&=&d^\dagger_{\sigma'}\left(\epsilon_0 \sigma_0+B\sigma_z\right)d_\sigma + U n_\uparrow n_\downarrow,  \nonumber\\
H_w&=& \int_0^{L_w} dx \,c^\dagger_{x\sigma'}\left[\left(\frac{\hbar^2k_x^2}{2m}-\mu\right)\sigma_0+\alpha k_x \sigma_y+ B\sigma_z\right]c_{x\sigma}  \nonumber\\
&&+ \Delta\left(c_{x\uparrow}c_{x\downarrow} + c^\dagger_{x\downarrow}c^\dagger_{x\uparrow}\right), \nonumber\\
H_\mathrm{hop} &=& t \left(c^\dagger_{0\sigma}d_{\sigma}+d^\dagger_{\sigma}c_{0\sigma}
\right),
\label{model}
\end{eqnarray}
where $\epsilon_0$ is the dot level, $U$ is its charging energy, $m$ is the nanowire's effective mass, $\Delta$ is the induced superconducting pairing, $\alpha$ is the spin-orbit coupling, $\mu$ is the nanowire's chemical potential and $B$ is a Zeeman splitting. In practical calculations, the nanowire is discretized into tight-binding sites at $a_0=10$ nm intervals. Operators $d_\sigma$ and $c_{x\sigma}$ denote electrons in the dot and (discrete) point $x$ of the nanowire, respectively, and $k_x=-i\partial_x$ is approximated by finite differences. Sums over spin indices $\sigma$ are implicit throughout this work. For $B>B_c\equiv\sqrt{\mu^{2}+\Delta^2}$ the nanowire enters a topological phase, with one Majorana state at each end, which we denote by $\gamma_L$ (inner or leftmost Majorana, close to the dot) and $\gamma_R$ (outer or rightmost Majorana, further from the dot), see Fig. \ref{fig:sketch}.

We are interested in the Coulomb blockade regime for the dot, where the relevant physics of a quantum dot coupled to a superconductor (singlet-doublet parity crossings) is well described within a self-consistent mean-field approximation of the interaction term in $H_d$, \begin{equation}
\label{meanfieldU}
U n_\uparrow n_\downarrow\approx U \left(n_\uparrow \langle n_\downarrow\rangle+\langle n_\uparrow\rangle n_\downarrow-\langle n_\uparrow\rangle\langle n_\downarrow\rangle\right).
\end{equation}
In principle, other pairing terms exist in the full mean-field decoupling due to the proximity of the superconductor. We have also performed calculations with the full mean field theory, including all possible decouplings, but the results in the low energy spectrum are almost indistinguishable from the above approximation using realistic parameters \footnote{It is well known that mean field solutions artificially break time reversal symmetry. However, we always focus on the large Zeeman regime, which is well described by a mean field approximation. Furthermore, we note that Kondo physics is only relevant in the regime $T_K/\Delta\gtrsim 0.6$ (see e.g. E. J. H. Lee {\it et al}, Phys. Rev. B {\bf 95}, 180502(R), 2017) which is far from the parameter regime we consider (with a doublet ground state for odd occupancy, i. e. at the center of the Coulomb Blockade diamond).}.

The contact between dot and nanowire is described using the simplest possible model, $H_\mathrm{hop}$. A more realistic alternative could be to model a smooth potential barrier between two segments of the wire created by a pinch-off gate underneath. In the case of a short dot and short barrier, however, the two models should be quantitatively similar for an proper choice of dot-nanowire hopping amplitude $t$ related to the barrier strength. The most important difference between the two contact models is that in the results to follow the information measured by the dot applies to some spatial average of the Majorana wave function inside the barrier, instead of simply at the endpoint of the decoupled wire. As an important remark, the dot-nanowire hopping amplitude $t$ modifies the position of the dot level $\epsilon_0$. In the following, $\epsilon_0$ will denote the actual dot level for a given $t$, taken in our simulations as 10\% of the hopping amplitude between neighboring sites in the nanowire (weak coupling limit).

\subsection{Spectral phenomenology of the tight-binding model}
\label{sec:TB}

\begin{figure}
   \centering
   \includegraphics[height=0.89\columnwidth]{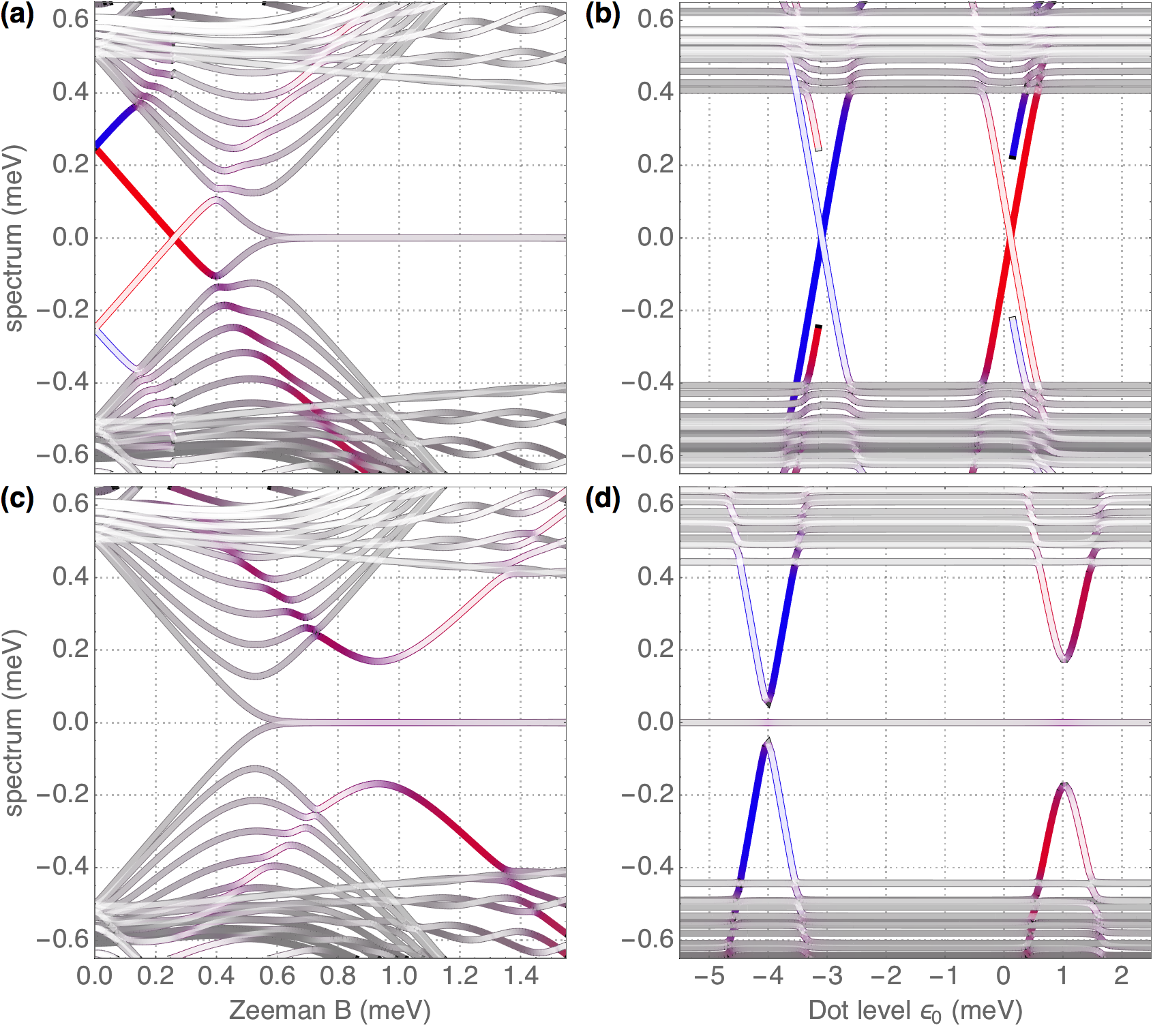}
   \caption{Low energy spectrum of the tight-binding dot-wire model as a function of $B$ for fixed $\epsilon_0=0.25$ meV (a) and $1.0$ meV (c), and as a function of $\epsilon_0$ for fixed $B=0.1$ meV (b) and $1.0$ meV (d).  Blue (red) lines correspond to spin-up (down) states concentrated on the dot, with electron-like (hole-like) character shown as solid (hollow) curves. Charge interactions in the dot are $U=3$ meV, spin-orbit coupling is $\alpha=60$ meV nm, and nanowire length is $L_w=2\mu$m. Panel (b) represents the paradigmatic singlet-doublet-singlet transitions in a dot-trivial superconductor junction with Zeeman-split dot levels, see Ref. \onlinecite{Lee:NN14}. Panel (d) is the analogue for a topological superconductor. While the dot states avoid the Majorana zero mode in (c,d), the zero Majorana energy is unperturbed by the dot-level crossing. This is a signature of the Majorana non-locality.}
   \label{fig:TBlong}
\end{figure}

\begin{figure}
   \centering
   \includegraphics[height=0.89\columnwidth]{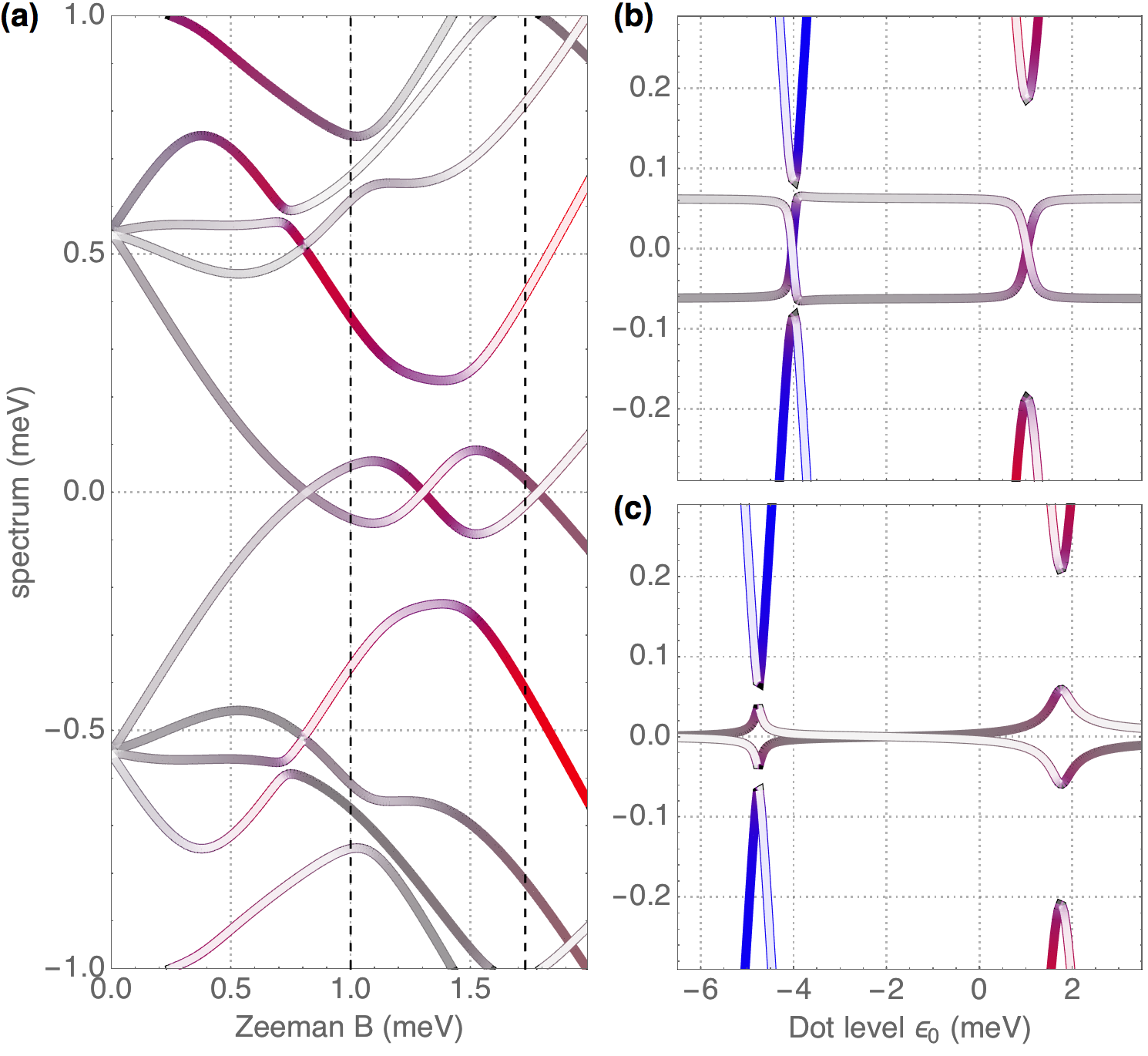}
   \caption{Same as Fig. \ref{fig:TBlong} but for a shorter $L_w=400$ nm. The Majoranas overlap, which leads to oscillatory splittings (a) and a finite hybridization with the dot states at the zero-energy crossings. This results in bowtie-like (b) or diamond-like (c) patterns in the Majorana energy as a function of dot level $\epsilon_0$. (a) assumes $\epsilon_0=1.25$, while (b) and (c) are taken at fixed $B=1.0$ and $1.7$ meV, respectively [vertical dashed lines in (a)].}
   \label{fig:TBshort}
\end{figure}

The spectrum of the dot-wire system, treating interactions within a self-consistent mean-field scheme, exhibits rather different subgap features depending on whether the nanowire is topological or not.
In the trivial phase $B<B_c$, only dot levels may appear below the superconducting gap. As expected, we recover the physics of singlet-doublet parity crossings:  subgap states, which are sometimes called Shiba states when they are spin-polarized by $B$, exhibit protected parity crossing at zero energy either as a function of Zeeman $B$ (for fixed  $\epsilon_0$ in the singlet regime) or as a function of $\epsilon_0$ for fixed $B$. 
 Fig. \ref{fig:TBlong}a shows these crossings in a long $L_w=2\mu$m InAs nanowire for $\mu=0$ and $\Delta=0.5$ meV and fixed $\epsilon_0\approx 0.25$ meV as a function of Zeeman energy $B$ (blue and red denote up and down spin polarizations along $z$, solid and hollow denote particle-hole character). Alternatively, Shiba states may cross zero energy twice (one per spin in the dot) for a fixed $B$ as $\epsilon_0$ is increased, the first at negative $\epsilon_0$ as the occupation of the dot goes from 2 to 1 (single-doublet), and the second at positive $\epsilon_0$ when it jumps from 1 to 0 (doublet-singlet). Fig. \ref{fig:TBlong}b shows these two crossings in the trivial phase for fixed $B=0.1$ meV.

In the topological phase $B>B_c$ of a sufficiently long nanowire, Majorana states arise at zero energy. Shiba states attempting to cross zero energy, Fig. \ref{fig:TBlong}c,d, are then forced into an anticrossing with Majorana states. Each of the two anticrossings has a different amplitude, but it is strictly finite for finite $t$.  For truly topological nanowires (much longer than the Majorana size), the two Majoranas do not overlap, which pins them to zero energy exactly, even across the resonance with the dot state \cite{Leijnse:PRB11}. In other words, while the dot state avoids crossing the Majorana zero mode, the Majorana itself is unperturbed by the resonance with the dot state. This is a direct manifestation of the Majorana non-locality, or of topological protection of the zero mode.

For shorter wires, comparable in length to the Majorana size, the two Majoranas overlap to some extent, which leads to an oscillatory splitting $\delta$ as a function of $B$ (and $\mu$), see Fig. \ref{fig:TBshort}a. In this case the crossing with the dot states does lead to a change in the energy of the split Majorana states. The crossings may then take the form of `bowtie'-like or `diamond'-like shapes for the split Majorana levels, Figs. \ref{fig:TBshort}b,c respectively. Intermediate patterns may also be observed (not shown). These type of spectral patterns are not uncommon in systems containing one or more Majorana nanowires of finite length \cite{Prada:PRB12,Vernek:PRB14,Ruiz-Tijerina:PRB15,Cayao:PRB15,Ricco:16,Sau:PRB17}.

From the above discussion it may already be anticipated that the sensitivity of overlapping Majoranas to the dot level could be exploited to detect their degree of non-locality and hence of topological protection. Moreover, a comparison between the strength of the two dot-Majorana resonances will be shown to directly probe the spin structure of the Majorana wave function itself. In the following we will derive a simple low energy description that will allow us to interpret the bowtie/diamond anticrossing structure, and show how it can be used to extract quantitative parameters of the Majorana wave function at the contact.

\section{Derivation of the effective low-energy model}
\label{sec:effmodel}

The goal is to derive an effective model that involves only the two electronic states in the dot and the two Majoranas, but that accurately describes the low energy sector of the full system. To this end we introduce four new parameters, that emerge from the full model: 
\begin{eqnarray*}
\delta &:& \textrm{splitting of overlapping Majoranas without the dot} \\
\theta_L &:& \textrm{spin canting angle of the left Majorana at $x=0$} \\
\theta_R &:& \textrm{spin canting angle of the right Majorana at $x=0$} \\
t_L &:& \textrm{hopping from the dot to the left Majorana $\gamma_L$}\\
t_R &:& \textrm{hopping from the dot to the right Majorana $\gamma_R$}
\end{eqnarray*}
We now discuss each of these in turn.

For a nanowire of finite length, the Majorana bound states $\gamma_L$ and $\gamma_R$ are not eigenstates of $H_w$. Due to their spatial overlap, they hybridize into a fermionic eigenstate $c_M^\dagger = (\gamma_L-i\gamma_R)/\sqrt{2}$ of energy $\delta$. The low energy effective Hamiltonian of the isolated nanowire in the topological phase thus reads
\[
H_w^\mathrm{eff} = \frac{\delta}{2} (c^\dagger_M c_M-c_M c^\dagger_M) = i\frac{\delta}{2}(\gamma_L\gamma_R-\gamma_R\gamma_L)= i\delta\,\gamma_L\gamma_R.
\]
The value of $\delta$ decays exponentially with $L_w$ and oscillates around zero with $k_FL_w$ \cite{Klinovaja:PRB12}. In the long wire limit, $\delta\to0$ and $\gamma_{L,R}$ become degenerate zero energy eigenstates. 

The Majorana bound states satisfy self-conjugation $\gamma_i=\gamma_i^\dagger$ and $\{\gamma_i,\gamma_j\}=\delta_{ij}$ ($i=L,R$). With this normalization, the $\gamma_i$ can therefore be thought of as two Bogoliubov quasiparticles with a special particle-hole conjugation symmetry.  Their spatial wave function structure in the continuum limit reads
\begin{equation}
\gamma_i=\frac{1}{\sqrt{2}}\int dx\, \left(u^{(i)}_{\sigma}(x)c^\dagger_{x\sigma}+u^{(i)*}_{\sigma}(x)c_{x\sigma}\right),
\end{equation}
with properly normalized $u^{(L,R)}_\sigma(x)$ and $v^{(L,R)}_\sigma(x)$, concentrated around the left and right ends of the nanowire. 
The outer (rightmost) Majorana is related to the inner (leftmost) Majorana by spatial and $\sigma_x$ inversion, see Fig. \ref{fig:sketch}b,
\begin{eqnarray}\label{u2}
u^{(R)}_{\uparrow}(x) &=& -iu^{(L)}_{\uparrow}(L_w-x), \nonumber\\
u^{(R)}_{\downarrow}(x) &=& iu^{(L)}_{\downarrow}(L_w-x).
\end{eqnarray}

For the isolated nanowire, the $u_\sigma^{(i)}(x)$ amplitudes vanish at $x=0$ and $x=L_w$. Close to the $x=0$ contact we may expand $u^{(i)}_{\sigma}(x)\approx x \,{u'_{\sigma}}^{(i)}(0)  +\mathcal{O}(x^2)$ \cite{Prada:EPJB04} and parametrize the slopes ${u'_{\sigma}}^{(i)}(0)$ by spin canting angles $\theta_{L,R}$ and real coefficients ${u'_0}^{(L,R)}$, 
\begin{eqnarray}\label{theta}
\left({u'_{\uparrow}}^{(L)}(0),{u'_{\downarrow}}^{(L)}(0)\right) &=& {u'_0}^{(L)}\left(\sin\frac{\theta_L}{2},-\cos\frac{\theta_L}{2}\right), \nonumber\\ 
\left({u'_{\uparrow}}^{(R)}(0),{u'_{\downarrow}}^{(R)}(0)\right) &=& -i {u'_0}^{(R)}\left(\sin\frac{\theta_R}{2},\cos\frac{\theta_R}{2}\right).
\end{eqnarray}
For highly non-local Majoranas, we have ${u'_0}^{(L)}\gg {u'_0}^{(R)}$.
The canting angle $\theta_L$ of the leftmost Majorana is independent of nanowire length $L_w$, but for the amplitude of the rightmost Majorana at $x=0$ the angle $\theta_R$
depends on $L_w$, see Fig. \ref{fig:sketch}b.
Both $\theta_{L,R}$ are moreover expected to depend on the Zeeman field $B$ and the spin-orbit coupling $\alpha$, as these two scales control the spin orientation of \emph{propagating} modes in the Rashba nanowire in the absence of superconductivity $\Delta$. The detailed relation is derived in Sec. \ref{sec:micro}. Note however, that as written above, the Majorana spinors lie in the $x-z$  plane of SU(2), while the effective Zeeman-Rashba field lies in the $y-z$ plane. This might appear surprising, but it is correct: the spin orientation of the electronic sector of Majorana bound states is orthogonal to the spin of propagating states \cite{Sticlet:PRL12}. In the limit of large B the Majorana spin orientation at the edge becomes polarized along $-B\hat{z}$ like that of propagating states ($\theta_L=0$), while corrections of order $\mathcal{O}(\Delta/B, \alpha/B)$ yield a spin canting $\theta_L>0$ along the nanowire direction $x$.

The coupling of the quantum dot and the nanowire in the low energy effective model distinguishes between hoppings to $\gamma_L$ and $\gamma_R$, see Fig. \ref{fig:sketch}. We may write 
\[
H_\mathrm{hop}^\mathrm{eff}=(t_{L\sigma} d^\dagger_\sigma-t^*_{L\sigma}d_\sigma)\gamma_L+(t_{R\sigma} d^\dagger_\sigma-t^*_{R\sigma}d_\sigma)\gamma_R.
\]
The hopping amplitudes $t_{i\sigma}$ arise from a wave function overlap between the dot states and the Majorana wave function at the $x=0$ edge, Eq. \eqref{theta}, so that $t_{L\sigma}=t a_0 {u'_\sigma}^{(L)} = t_L (\sin\frac{\theta_L}{2},-\cos\frac{\theta_L}{2})$ (larger hoppings) and $t_{R\sigma}=t a_0{u'_\sigma}^{(R)}  = -i t_R (\sin\frac{\theta_R}{2},\cos\frac{\theta_R}{2})$ (usually smaller than $t_{L\sigma}$, exponentially suppressed with increasing $L_w$). Here we have defined the real hopping amplitudes $t_i \equiv t  a_0{u'_0}^{(i)}$ from the dot to each of the two Majoranas, with   $a_0$ the tight-binding lattice spacing  \footnote{In the continuum limit the Majorana wave function vanishes at the ends of the nanowire, $u_\sigma^{(L,R)}(0)=0$. In that case, the hopping amplitudes $t_{L\sigma,R\sigma}$ are proportional to the spatial derivatives $\partial_x u_\sigma^{(L,R)}(x=0)$ times a lengthscale associated to the contact \cite{Prada:EPJB04}. In our minimal contact model within a discretized nanowire tight-binding description, such lengthscale is the lattice constant $a_0$. With this choice, the hopping amplitudes becomes proportional to the Majorana amplitudes at the leftmost site of the discretised nanowire, which are no longer zero, but $a_0 {u'_0}^{(L,R)}(0)$.}.

The last piece of our low energy model is the effective Hamiltonian for the dot, which we take just as $H_d$ in Eq. (\ref{model}), with the same mean-field decoupling of Eq. \eqref{meanfieldU}. In this case, however, the mean field self-consistency is approximated by the analytical mean field solution for a decoupled dot, which for $B>0$ reads
\begin{eqnarray}\label{meanfield}
\langle n_\downarrow\rangle&=&\langle n_\uparrow\rangle = 1 \space \textrm{ for $\epsilon_0<-U-B$}, \nonumber\\
\langle n_\downarrow\rangle&=&1-\langle n_\uparrow\rangle = 1\textrm{ for $-U-B<\epsilon_0<B$}, \nonumber \\
\langle n_\downarrow\rangle&=&\langle n_\uparrow\rangle = 0 \space \textrm{ for $B<\epsilon_0$}.
\end{eqnarray}

The complete effective model then reads, up to a constant $-U\langle n_\uparrow\rangle\langle n_\downarrow\rangle$, as
\begin{widetext}
\begin{eqnarray}
H^\mathrm{eff}&=&\frac{1}{2}\left(d^\dagger_\uparrow,d^\dagger_\downarrow,d_\uparrow,d_\downarrow,\gamma_L,\gamma_R\right)\check{H}^\mathrm{eff}\left(d_\uparrow,d_\downarrow,d^\dagger_\uparrow,d^\dagger_\downarrow,\gamma_L,\gamma_R\right)^T, \nonumber\\
\frac{1}{2}\check{H}^\mathrm{eff}&=&\left(\begin{array}{cccccc}
\frac{\epsilon_0+B+U\langle n_\downarrow\rangle}{2} & 0 & 0 & 0  &  t_L\sin\frac{\theta_L}{2} & -i t_R \sin\frac{\theta_R}{2}\\
0 & \frac{\epsilon_0-B+U\langle n_\uparrow\rangle}{2}   & 0 & 0  & -t_L\cos\frac{\theta_L}{2} & -i t_R \cos\frac{\theta_R}{2}\\
0 & 0 & -\frac{\epsilon_0+B+U\langle n_\downarrow\rangle}{2} & 0 & -t_L\sin\frac{\theta_L}{2} & -i t_R \sin\frac{\theta_R}{2}\\
0 & 0 & 0 & -\frac{\epsilon_0-B+U\langle n_\uparrow\rangle}{2}   &  t_L\cos\frac{\theta_L}{2} & -i t_R \cos\frac{\theta_R}{2}\\
t_L\sin\frac{\theta_L}{2} & -t_L\cos\frac{\theta_L}{2} & -t_L\sin\frac{\theta_L}{2} & t_L\cos\frac{\theta_L}{2}     & 0  &  i\delta/2 \\
it_R\sin\frac{\theta_R}{2} & it_R\cos\frac{\theta_R}{2} & it_R\sin\frac{\theta_R}{2} & it_R\cos\frac{\theta_R}{2} & -i\delta/2 & 0
\end{array}\right).
\label{effmodel}
\end{eqnarray}
\end{widetext}
Due to our choice of normalization for $\gamma_{L,R}$, the spectrum of the system is simply given by the eigenvalues of the above $\check{H}^\mathrm{eff}$ matrix.

\begin{figure*}
   \centering
   \includegraphics[width=\textwidth]{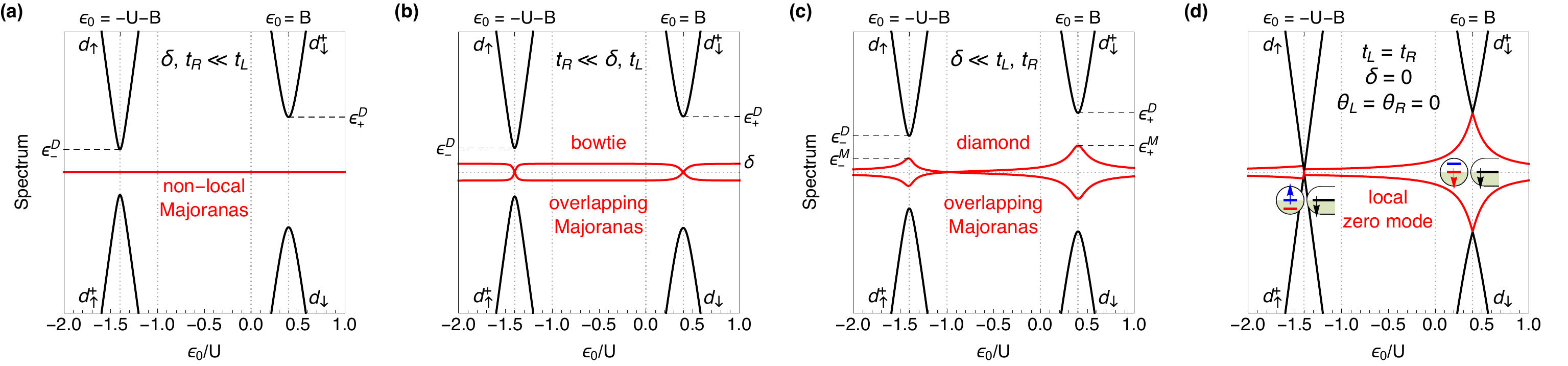}
   \caption{Low energy spectrum, using the effective model, of the quantum dot coupled to the Majorana nanowire as a function of dot level $\epsilon_0$. In panel (a) we show  the limit of two decoupled Majoranas, $\delta,t_R\to 0$ (sufficiently long nanowire), whose energy remains pinned to zero across the Shiba-Majorana resonances. Panel (b) shows the bowtie-like case, for which the Majorana splitting $\delta$ is finite, but the direct hopping $t_R$ from the dot to the rightmost Majorana is suppressed. Panel (c) corresponds to the diamond-like case, where such direct hopping is not suppressed, and the Majorana splitting $\delta$ is tuned close to zero. Patterns similar to (a,c) were recently observed by M. Deng et al. \cite{Deng:inprep}. The trivial case of a strictly local zero mode relevant for very short wires (Shiba state with energy $\delta$ tuned to zero) is shown in (d). The complete asymmetry between crossings reveals complete lack of spin canting ($\theta_L=\theta_R=0$), while the lack of splitting between dot and Majorana lines (black and red) reveals complete locality ($t_L=t_R$). }
   \label{fig:levels}
\end{figure*}

\begin{figure}
   \centering
      \includegraphics[width=1\columnwidth]{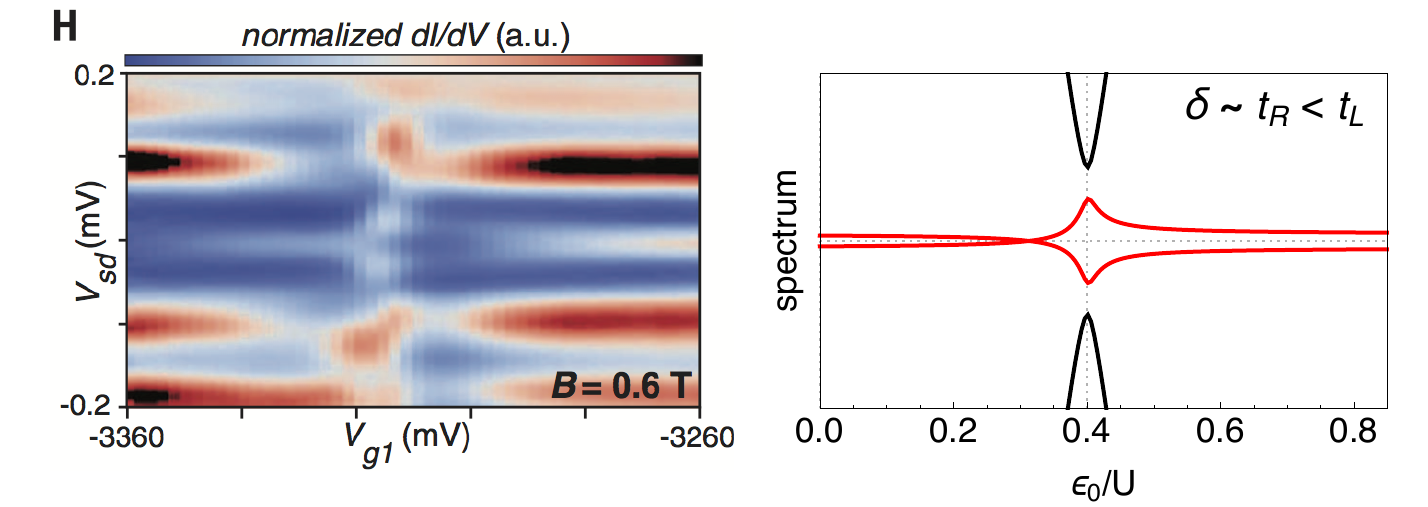}
   \caption{(a) Asymmetric Shiba-Majorana anticrossing measured in the group of C. Marcus \cite{Deng:S16}, reproduced with permission. (b) Similar structure obtained using the effective model of Eq. \eqref{effmodel} in the $\delta\sim t_R<t_L$ case.}
   \label{fig:measurement}
\end{figure}

\subsection{Interpretation of dot-Majorana anticrossings using the effective model}
\label{sec:int}

The spectrum of the effective model as a function of dot level $\epsilon_0$, shown in Fig. \ref{fig:levels}, replicates the phenomenology of the full tight-binding model. It is characterized by the two parity crossings in the dot, at $\epsilon_0=-U-|B|$ and $\epsilon_0=|B|$, where the average dot occupation changes according to Eq. \eqref{meanfield}. At these two points, the spectrum has four levels close to zero energy, two in the dot and two (Majorana) in the wire, which then anticross. The fundamental difference between the $\epsilon_0<0$ and $\epsilon_0>0$ anticrossings is the spin polarization of the low energy dot excitations involved, which is opposite ($d^\dagger_\uparrow, d_\uparrow$ on the negative and $d^\dagger_\downarrow, d_\downarrow$ on the positive crossing, see top and bottom dot configurations in Fig. \ref{fig:sketch}b, respectively). Thus, if the spin canting of the Majorana is not very large (small $\theta_L$), the negative and positive resonances results in quite \emph{different} anticrossing strengths of the four states. The $\epsilon_0<0$ anticrossing tends to be smaller for realistic nanowire parameters, since the Majorana spin has a larger $\downarrow$ component ($0<\theta_L<\pi/2$) and is thus more orthogonal to the $\uparrow$ dot excitations. Conversely, in the $\epsilon_0>0$ crossing, low-energy dot excitations have a $\downarrow$ spin, mostly parallel to the Majorana's, and hybridize more strongly. Strong canting thus translates into almost symmetric crossings, and weak canting (Majorana spin at the end of the nanowire mostly polarized along Zeeman field) into strongly asymmetric crossings.

The shape of each anticrossings (bowtie, diamond or something in between) is controlled by the relative value of two energy scales that are both much smaller than the others, at least for nanowires in the hundreds of nanometers: the Majorana splitting $\delta$ and the dot-outer Majorana hopping $t_R$. For Majoranas with some degree of non-locality, the hopping $t_L$ is larger than $t_R$, as the inner Majorana is more strongly coupled to the dot. The splitting $\delta$ is oscillatory with nanowire parameters, such as $L_w$, $B$ or $\mu$, so it can become zero at specific points in parameter space. $t_R$ in contrast also oscillates but remains finite.

Figure \ref{fig:levels} presents four typical spectra for the effective dot-nanowire model of Eq. \eqref{effmodel}. This time, black lines correspond to levels predominantly in the quantum dot, and red to Majorana-like levels in the nanowire. The Majorana M and dot D energy levels at the $\epsilon_0=-U-|B|$ ($-$) and $\epsilon_0=|B|$ ($+$) anticrossings are denoted by $\epsilon_{\pm}^{M,D}\geq 0$, respectively. Panel (a) shows the case of non-overlapping Majoranas, with $\delta=t_R=0$. This is the situation of truly non-local, topologically protected Majorana zero modes. Panel (b) shows a case with a finite $\delta\gg t_R$. In this case, the two anticrossings exhibit a symmetric bowtie structure of width $2\delta$ for Majorana levels ($\epsilon_{\pm}^{M}=0$ at both anticrossings), and an asymmetric anticrossing for dot levels, with different amplitudes $\epsilon_{-}^{D}<\epsilon_{+}^{D}$ in general. As discussed, the degree of symmetry between dot-level anticrossings is directly related to the amount of spin canting of the inner Majorana,  $\theta_L=\pi/4$ in this simulation.
Panel (c) shows the opposite situation, wherein $t_R\gg \delta$.  We see that the anticrossing structure is very different in this case, with an asymmetric diamond-like shape for the Majorana levels ($\epsilon_{+}^{M}>\epsilon_{-}^{M}>0$). 
The degree of symmetry between the two diamond heights is related to the amount of spin canting $\theta_R$ of the \emph{outer} Majorana, see below. 
Anticrossings similar to Figs. 
\ref{fig:levels}(a,c) were recently measured in the group of C. Marcus \cite{Deng:inprep}. More generic anticrossings with a skewed profile may also arise and have also been observed, see e.g. Fig.
\ref{fig:measurement}. 

The limiting case of a strictly local zero mode (trivial Shiba state tuned to zero energy in a very short nanowire) corresponds to $\delta=0$, $t_L=t_R$ and $\theta_L=\theta_R=0$ (no canting, full asymmetry between the two dot resonances), which results in the characteristic anticrossing pattern shown in Fig. \ref{fig:levels}d, wherein diamonds are not spectrally separated from dot levels. This lack of spectral separation is a generic feature of $t_L=t_R$ and occurs for any $\delta$ and $\theta_{L,R}$. It can therefore be used as a simple method to diagnose local subgap states.

All the important parameters in the model ($\delta,t_{L,R},\theta_{L,R}$) can be quantified from the structure of the two anticrossings. To make the connection clearer, we derive general analytical expressions for the four positive energy levels $\epsilon_{\pm}^{M,D}$ at resonance, valid for large $B$ but for any values of $\delta$ and $t_{L,R}$. They are obtained by projecting out the high-energy dot levels not involved at each anticrossing, and explicitly diagonalizing the resulting $4\times4$ Hamiltonian in this low-energy subspace. We obtain
\begin{eqnarray*}
\epsilon_-^{M,D}&=&\sqrt{\frac{\delta^2}{2}+s_L^2+s_R^2\mp\sqrt{\left(\frac{\delta^2}{2}+s_L^2+s_R^2\right)^2-4s_L^2s_R^2}},\\
\epsilon_+^{M,D}&=&\sqrt{\frac{\delta^2}{2}+c_L^2+c_R^2\mp\sqrt{\left(\frac{\delta^2}{2}+c_L^2+c_R^2\right)^2-4c_L^2c_R^2}},
\end{eqnarray*}
where $s_i=2t_i\sin\frac{\theta_i}{2}$ and $c_i=2t_i\cos\frac{\theta_i}{2}$. 

In the case of $\delta, t_R\ll t_L$ (decoupled Majoranas, Fig. \ref{fig:levels}a), $\epsilon_\pm^{M}=0$ and $\epsilon_{\pm}^{D}=2t_L\sqrt{1\pm\cos\theta_L}$. The limit $t_R\ll\delta,t_L$ also yields $\epsilon_{\pm}^{M}=0$ and $\epsilon_{\pm}^{D}=\sqrt{\delta^2+4t_L^2\left(1\pm\cos\theta_L\right)}$, see Fig. \ref{fig:levels}b. In the opposite case $\delta\ll t_L,t_R$ of Fig. \ref{fig:levels}c we have $\epsilon_{\pm}^{M}=2t_R\sqrt{1\pm\cos\theta_R}$ and $\epsilon_{\pm}^{D}=2t_L\sqrt{1\pm\cos\theta_L}$. Note that in all cases the dot-like energies $\epsilon_{\pm}^{D}$ depend only on the inner-Majorana canting $\theta_L$ and hopping $t_L$, and the Majorana-like energies $\epsilon_{\pm}^{M}$ contain information about the outer Majorana canting $\theta_R$ and hopping $t_R$.

\begin{figure*}
   \centering
      \includegraphics[width=\textwidth]{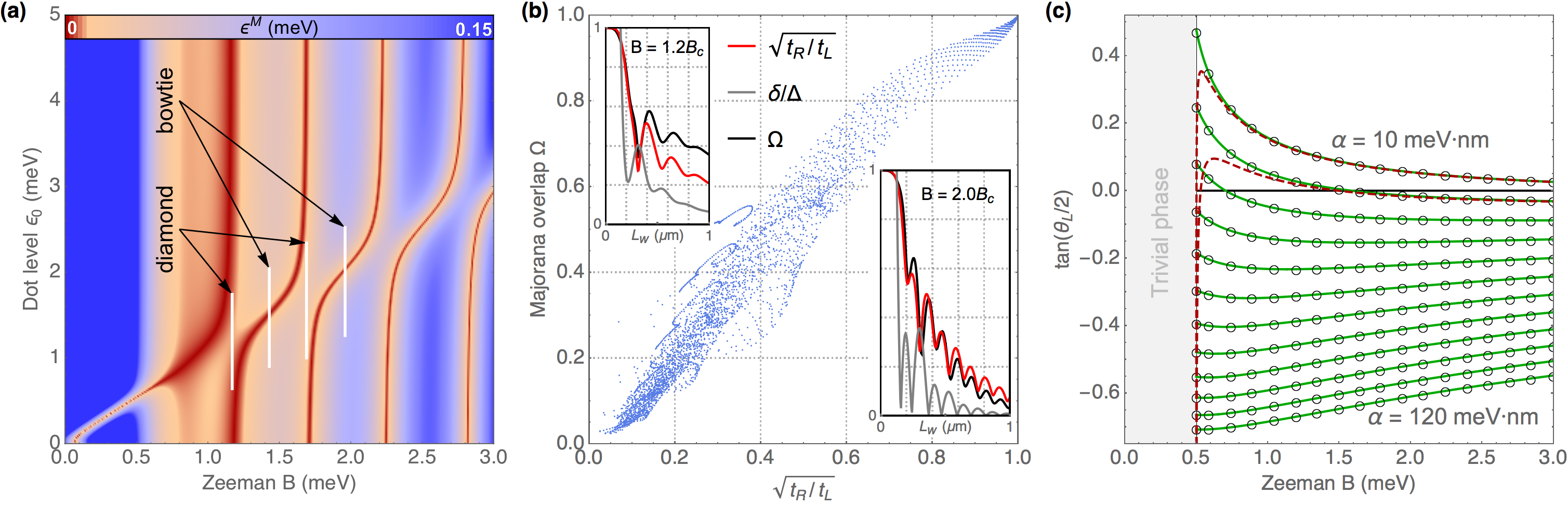}
   \caption{(a) Energy $\epsilon^M$ of the lowest energy state, corresponding to the hybridized Majoranas in a $L_w=1\mu$m nanowire coupled to a quantum dot. As a function of Zeeman $B$, this splitting becomes zero repeatedly (red) as a result of the oscillatory behaviour of $\delta$. At dot-Majorana crossings, $\epsilon_0=|B|$ (shown) and $\epsilon_0=-U-|B|$ (not shown), the vertical $\epsilon^M=0$ lines become deflected. Different constant-$B$ cuts across the dot-Majorana resonance can exhibit bowtie or diamond shapes, see vertical white lines. (b) Correlation between the degree of Majorana non-locality $\Omega$, Eq. \eqref{omega}, and the local estimator $\sqrt{t_R/t_L}$, both computed within the tight-binding model of Eq. \eqref{model} for different values of $B>B_c$ and $L_w$ at $\mu=0$ (other realistic values of $\mu$ yield similar results). Insets show the dependence of $\Omega$, $\sqrt{t_R/t_L}$ and $\delta/\Delta$ with $L_w$ for $B$ just above the critical $B_c$  and at higher $B$  \cite{Klinovaja:PRB12}. (c) Spin-canting $\tan(\theta_L/2)$ of the left Majorana as a function of $B$ for $\Delta=0.5$ meV and $\mu=0$ as spin-orbit coupling $\alpha$ increases in steps of $10$ meV nm. Circles correspond to tight-binding, solid lines to Eq. \eqref{thetaL}, and dashed lines to the weak spin-orbit approximation of Eq. \eqref{weakso}.}
   \label{fig:last}
\end{figure*}

We can derive useful quantitative information from the ratios of the four energies. The ratio between the dot-like energies $\epsilon^D_\pm$ in the two anticrossings directly yields the inner Majorana canting $\theta_L$ if $\delta\ll t_L$ for any $t_R$,
\begin{equation}\label{thetaLratio}
\frac{\epsilon_{-}^{D}}{\epsilon_{+}^{D}}=\left|\tan\frac{\theta_L}{2}\right|  \textrm{\space (for $\delta\ll t_L$)}.
\end{equation}
Similarly, a comparison of $\epsilon^M_\pm$ at $\delta\ll t_R$ (i.e. at diamond-like anticrossings) yields the outer Majorana canting $\theta_R$ at the contact, 
\begin{equation}\label{thetaRratio}
\frac{\epsilon_{-}^{M}}{\epsilon_{+}^{M}}=\left|\tan\frac{\theta_R}{2}\right|  \textrm{\space (for $\delta\ll t_L,t_R$)}.
\end{equation}
Once $\theta_L$ and $\theta_R$ are known using the above, we can measure the ratio from the dot and Majorana energies within each anticrossing to extract the ratio $t_R/t_L$, 
\begin{eqnarray}\label{tLtRratio1}
\frac{\epsilon_{-}^{M}}{\epsilon_{-}^{D}}&=&\left|\frac{t_R}{t_L}\frac{\sin\frac{\theta_R}{2}}{\sin\frac{\theta_L}{2}}\right|  \textrm{\space (for $\delta\ll t_L,t_R$)},\\
\frac{\epsilon_{+}^{M}}{\epsilon_{+}^{D}}&=&\left|\frac{t_R}{t_L}\frac{\cos\frac{\theta_R}{2}}{\cos\frac{\theta_L}{2}}\right|  \textrm{\space (for $\delta\ll t_L,t_R$)}.\label{tLtRratio2}
\end{eqnarray}
This fully characterizes analytically the energies at the two anticrossings in terms of physical properties of the Majorana wave function. 

The energy of the lowest level $\epsilon^M$ for a general $B$ and $\epsilon_0$ is shown in Fig. \ref{fig:last}a for a $1\mu$m nanowire using the full tight-binding model. This plot allows one to understand the evolution of the $\epsilon^M=0$ parity crossing (red) throughout the $\epsilon_0,B$ plane, and the alternation of bowtie and diamond anticrossings as $B$ increases. In particular, bowtie (diamond) anticrossings as a function of $\epsilon_0$ appear at values of $B$ with maximum $\delta$ ($\delta=0$), see solid white lines. Other values of $B$ exhibit a varying amount of anticrossing asymmetry, see e.g. Fig. \ref{fig:measurement}b.

\section{Quantifying non-locality with local measurements}
\label{sec:nonlocal}

The degree of non-locality of the two Majoranas is defined by their overlap 
\begin{eqnarray}\label{omega}
\Omega&=&\sum_\sigma\int_0^{L_w} dx\left|u^{(L)}_\sigma(x)u^{(R)}_\sigma(x)\right|.
\end{eqnarray}
This quantity ranges from $0$ (no overlap) to $1$ (perfect overlap) and is connected to the charge of overlapping Majoranas by $Q=e\Omega$ \cite{Ben-Shach:PRB15, Dominguez:NQM17}. It is also related to the resilience of Majorana qubit to local environmental noise, with complete non-locality $\Omega=0$ signalling topological qubit protection. The value of $\Omega$ is sometimes incorrectly identified with the Majorana splitting $\delta/\Delta$, arguably because $\Omega=0$ implies $\delta=0$. However, the converse is not true. It is known than $\delta=|\langle \gamma_L|H_w|\gamma_R\rangle|$ can vanish at special points (parity crossings \cite{Lim:PRB12, Prada:PRB12, Das-Sarma:PRB12,Rainis:PRB13}) or extended parameter regions (pinned Majoranas \cite{Dominguez:NQM17}) even when Majoranas overlap, while $\Omega$ by definition cannot. In this section we show that, in contrast to $\delta/\Delta$, the ratio $t_R/t_L=|\langle x=0|\gamma_R\rangle/\langle x=0|\gamma_L\rangle|$ is an accurate estimator of $\Omega$, despite being a purely local quantity. More specifically
\begin{equation}
\Omega\approx \sqrt{t_R/t_L}.
\end{equation}
It is easy to see that both in the strictly local case $\Omega=1$ and in the completely non-local case $\Omega=0$, the above relation holds. To evaluate it for intermediate overlaps, we have computed the Majorana wave functions of depleted nanowires described by $H_w$ in Eq. \eqref{model} for a range of lengths $L_w$ and Zeeman fields $B>B_c$, extracting $\Omega$ and $\sqrt{t_R/t_L}$ for each. Figure \ref{fig:last}b shows a plot of the two quantities for all simulations, which shows a high correlation, with an correlation coefficient exceeding $0.95$. In the insets, we show $\Omega$, $\sqrt{t_R/t_L}$ and $\delta/\Delta$ as a function of nanowire length $L_w$. We see that for $B$ just above the critical $B_c$ (left inset), a regime in which the Majorana bound states in our model exhibit a double-exponential decay \cite{Klinovaja:PRB12}, $\sqrt{t_R/t_L}$ underestimates the overlap $\Omega$. At higher magnetic fields (right inset) only one decay lengthscale survives, and $\sqrt{t_R/t_L}$ becomes an essentially exact estimator of $\Omega$.

\subsection{Overlap estimator beyond uniform nanowires}
\label{sec:beyond}

\begin{figure}[t!]
   \centering
   \includegraphics[width=\columnwidth]{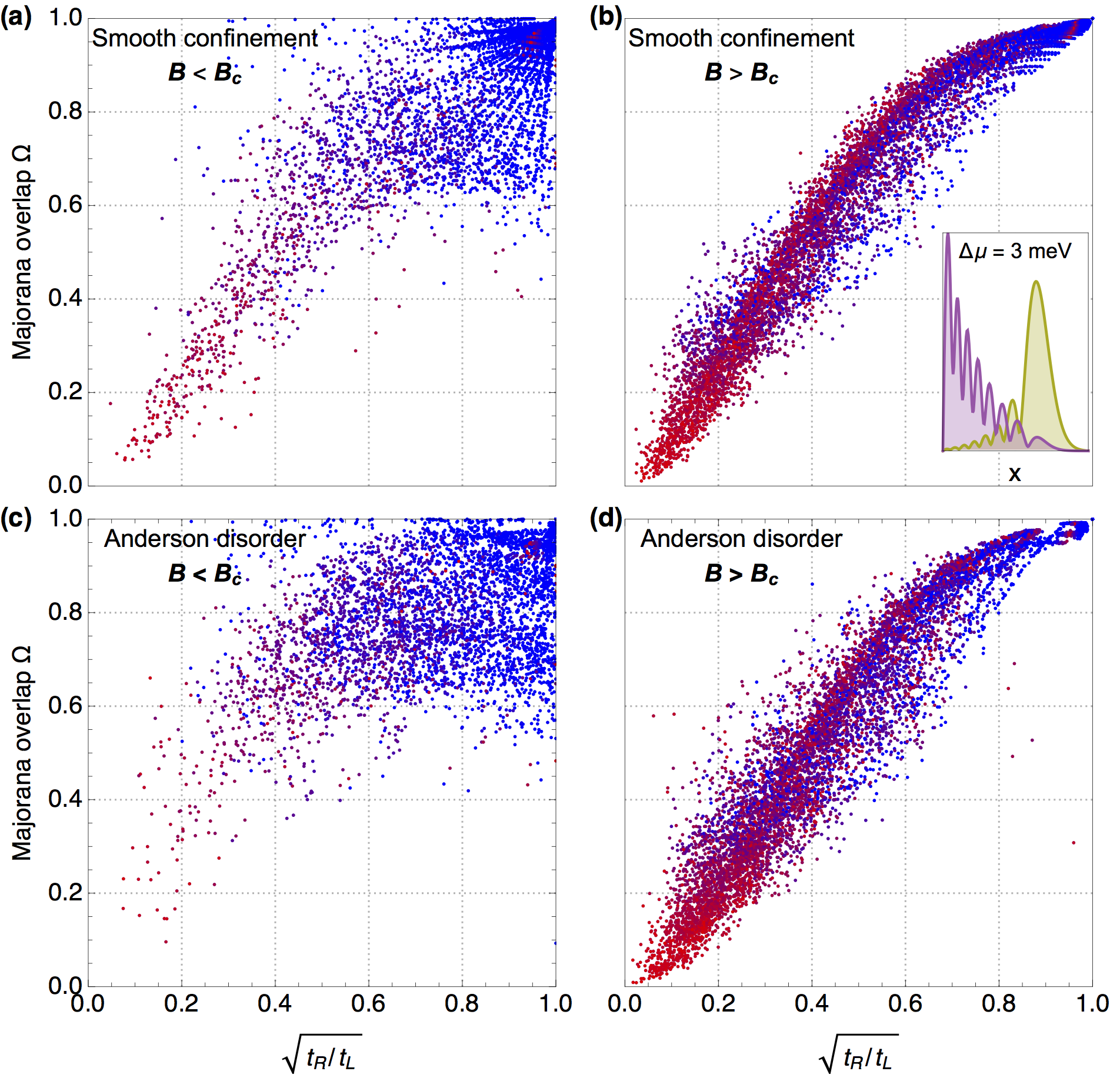}
   \caption{Correlation plots between $\sqrt{t_R/t_L}$ and the Majorana overlap $\Omega$, defined from the lowest Andreev bound state. Each dot corresponds to a different choice of system parameters. Dot color encodes the Majorana hybridization $\delta/\Delta$, from zero (red) to one (blue). Panels (a,b) correspond to nanowires ($\Delta=0.5$ meV, $\alpha=60$ meV nm) with a smoothly varying Fermi energy $\mu(x)=\mu_0+\Delta\mu x/L_w$. We vary $\Delta\mu=0\textrm{ to }3$ meV, $L_w=5\textrm{ to }1000$ nm, $\mu_0=-1.5\textrm{ to }1.5$ meV and $B=0\textrm{ to }B_c$ (a) or $B=B_c\textrm{ to }3B_c$ (b), where $B_c=\sqrt{\mu_0^2+\Delta^2}$. Panels (c,d) show the correlation in the presence of Anderson disorder, $\mu(x)=\mu_0+V(x)$, with $V(x)$ a spatially uncorrelated random potential uniformly distributed between $-\Delta\mu$ and $\Delta\mu$, and with $\Delta\mu$ again ranging from $0$ to 3 meV.}
   \label{fig:correlators}
\end{figure}

The above analysis connecting the structure of dot-Majorana anticrossings, a strictly local measurement, to the degree of Majorana overlap is made possible by a stringent assumption about the form of the Majorana wavefunction along the nanowire. Indeed, by comparing the estimator $\sqrt{t_R/t_L}$ to the overlap $\Omega$ for different parameter values of the $H_w$ model in Eq. \eqref{model}, we are effectively assuming that the Majorana wavefunction should always be of the Oreg-Lutchyn type \cite{Klinovaja:PRB12} for pristine, uniform nanowires. While this is a frequent assumption in the literature, it might be incomplete or not even apply in real samples. One should therefore consider whether physically sound generalizations of the Oreg-Lutchyn model could break the relation between $\sqrt{t_R/t_L}$ and $\Omega$.

We have analysed several such extensions, including extended quantum dots, disorder and screened potentials in the nanowires. We have found that even in these cases, $\sqrt{t_R/t_L}$ remains, perhaps surprisingly, a rather faithful estimator of $\Omega$ throughout the (nominally) topological regime. To back this claim we now present simulations for inhomogeneous and disordered nanowires.

Figure \ref{fig:correlators}(a,b) shows specifically the correlation between $\sqrt{t_R/t_L}$ and $
\Omega$ for extended models of $H_w$ with a non-uniform but smooth $\mu(x)=\mu_0+\Delta\mu (x/L_w)$ (with $\Delta\mu>0$ physically produced e.g. by inhomogeneous screening from the environment or non-uniform charge transfer from the parent superconductor). Panel (a) shows a result analogous to Fig. \ref{fig:last}b, for lowest-lying states in nanowires that are trivial at all points [i.e. $B<B_c(x=0)=\sqrt{\mu_0^2+\Delta^2}<B_c(x)$], while panel (b) corresponds to nanowires for which a finite portion around $x=0$ is non-trivial $B>B_c(x=0)$. Each point corresponds to a different set of values for $B$, $\mu$, $\Delta\mu$ and $L_w$. In these plots, the color of each point encodes the normalized energy $\delta/\Delta$ of the lowest-lying state, ranging from zero (red) to one (blue). As in Fig. \ref{fig:last}b, the estimator $\sqrt{t_R/t_L}$ and $\Omega$ remain highly correlated in this generalized model, with an correlation coefficient still exceeding 95\% in the non-trivial regime, panel (b). This is despite the fact that the Majorana wavefunctions with this smooth $\mu(x)$ confinment strongly deviates from the uniform case [particularly for $\gamma_R$, see inset to panel (b)].

A similar computation was carried out for $H_w$ in the presence of Anderson disorder $\mu(x)=\mu_0+V(x)$, where $V(x)$ is a spatially uncorrelated random potential uniformly distributed between $-\Delta\mu$ and $\Delta \mu$. Figure \ref{fig:correlators}(c,d) shows the correlation of the estimator in this case, for $\Delta \mu$ again ranging from zero to 3 meV, and with a different disorder realization for each choice of the rest of parameters. Again the overlap estimator remains very faithful in the non-trivial regime, panel (d), with a correlation around 95\%.

We thus conclude that a direct measurement of the quantity $\sqrt{t_R/t_L}$ via dot-Majorana anticrossings provides an accurate estimate of the degree of Majorana non-locality under rather general conditions, even beyond uniform nanowire models.

\section{Conclusions}
\label{sec:conclusions}

The field of topological superconductivity has reached a stage in which one may cautiously claim that Majorana zero modes have finally been detected experimentally. An important next step is now to carefully characterize these states, and in particular to demonstrate their unique non-local nature, which is the key to their promise as topologically protected logic elements for quantum computation. In this work we have shown that Majorana non-locality may be demonstrated through transport spectroscopy across currently available quantum dot-nanowire junctions. The key is to analyse the changes in the energy of Majorana states as they resonate with the  quantum dot level around zero energy. Only a true non-local Majorana completely decoupled from its partner will remain pinned to zero energy across the resonance, while the dot state anticrosses. The combination of an insensitive zero mode plus a dot-state anticrossing constitutes an essentially unambiguous signature of Majorana non-locality, assuming a rather general model for the dot-nanowire system. Deviations from strict non-locality become visible in the form of bowtie and diamond-like lineshapes of the Majoranas across the resonance. These may be used to quantify the degree of non-locality $\Omega$, in particular by the estimator $\Omega\approx\sqrt{t_R/t_L}$, as extracted e.g. from  diamond lineshapes using Eqs. (\ref{thetaLratio}-\ref{tLtRratio2}). This estimator remains highly accurate in the topological regime even in the presence of Anderson disorder or non-uniform potentials in the nanowire.

Furthermore, the spin polarization of the dot state, a result of single occupancy in the Coulomb blockade regime, also allows the dot to probe the internal spin structure of the Majorana zero mode at the end of the nanowire, i.e. its degree of spin canting. Moreover, as the orientation of the Majorana electron-spin depends on the nanowire's spin-orbit coupling, it becomes possible to quantitatively measure the latter, with the aid of simple analytical formulas, by comparing two consecutive dot anticrossings as the Zeeman field is increased. 

In summary, and rather remarkably, a local junction to a quantum dot is found to be capable of extracting the most relevant properties of the Majorana wave function, to quantify the degree of its topological protection, and to distinguish true non-local Majorana zero modes from other forms of zero modes, such as e.g. overlapping Majoranas subject to electrostatic pinning \cite{Dominguez:NQM17}, or topologically trivial parity crossings of non-topological origin \cite{Lee:PRL12,Lee:NN14,Jellinggaard:PRB16}. Implications of our findings for quantum information measurement-only protocols based on quantum dots coupled to Majoranas \cite{Karzig:16,Gharavi:PRB16,Hoffman:PRB16}, should be the subject of a future work.

\emph{Note added:} While finalising this manuscript a preprint was posted online \cite{Clarke:17} which partially overlaps with some of our results regarding Majorana non-locality detection, although in a simpler spinless setting.

\acknowledgements
We wish to thank K. Flensberg, C. Marcus and M. Deng for illuminating discussions. We acknowledge financial support from the Spanish Ministry of Economy and Competitiveness through Grant Nos. FIS2015-65706-P, FIS2015-64654-P and FIS2016-80434-P (AEI/FEDER, EU), the Ram\'on y Cajal programme, Grant Nos. RYC-2011-09345 and RYC-2013-14645 and the ``Mar\'ia de Maeztu'' Programme for Units of Excellence in R\&D (MDM-2014-0377)

\bibliography{biblio}

\begin{thebibliography}{58}%
\makeatletter
\providecommand \@ifxundefined [1]{%
 \@ifx{#1\undefined}
}%
\providecommand \@ifnum [1]{%
 \ifnum #1\expandafter \@firstoftwo
 \else \expandafter \@secondoftwo
 \fi
}%
\providecommand \@ifx [1]{%
 \ifx #1\expandafter \@firstoftwo
 \else \expandafter \@secondoftwo
 \fi
}%
\providecommand \natexlab [1]{#1}%
\providecommand \enquote  [1]{``#1''}%
\providecommand \bibnamefont  [1]{#1}%
\providecommand \bibfnamefont [1]{#1}%
\providecommand \citenamefont [1]{#1}%
\providecommand \href@noop [0]{\@secondoftwo}%
\providecommand \href [0]{\begingroup \@sanitize@url \@href}%
\providecommand \@href[1]{\@@startlink{#1}\@@href}%
\providecommand \@@href[1]{\endgroup#1\@@endlink}%
\providecommand \@sanitize@url [0]{\catcode `\\12\catcode `\$12\catcode
  `\&12\catcode `\#12\catcode `\^12\catcode `\_12\catcode `\%12\relax}%
\providecommand \@@startlink[1]{}%
\providecommand \@@endlink[0]{}%
\providecommand \url  [0]{\begingroup\@sanitize@url \@url }%
\providecommand \@url [1]{\endgroup\@href {#1}{\urlprefix }}%
\providecommand \urlprefix  [0]{URL }%
\providecommand \Eprint [0]{\href }%
\providecommand \doibase [0]{http://dx.doi.org/}%
\providecommand \selectlanguage [0]{\@gobble}%
\providecommand \bibinfo  [0]{\@secondoftwo}%
\providecommand \bibfield  [0]{\@secondoftwo}%
\providecommand \translation [1]{[#1]}%
\providecommand \BibitemOpen [0]{}%
\providecommand \bibitemStop [0]{}%
\providecommand \bibitemNoStop [0]{.\EOS\space}%
\providecommand \EOS [0]{\spacefactor3000\relax}%
\providecommand \BibitemShut  [1]{\csname bibitem#1\endcsname}%
\let\auto@bib@innerbib\@empty
\bibitem [{\citenamefont {Nayak}\ \emph {et~al.}(2008)\citenamefont {Nayak},
  \citenamefont {Simon}, \citenamefont {Stern}, \citenamefont {Freedman},\ and\
  \citenamefont {Das~Sarma}}]{Nayak:RMP08}%
  \BibitemOpen
  \bibfield  {author} {\bibinfo {author} {\bibfnamefont {C.}~\bibnamefont
  {Nayak}}, \bibinfo {author} {\bibfnamefont {S.}~\bibnamefont {Simon}},
  \bibinfo {author} {\bibfnamefont {A.}~\bibnamefont {Stern}}, \bibinfo
  {author} {\bibfnamefont {M.}~\bibnamefont {Freedman}}, \ and\ \bibinfo
  {author} {\bibfnamefont {S.}~\bibnamefont {Das~Sarma}},\ }\href@noop {}
  {\bibfield  {journal} {\bibinfo  {journal} {Rev. Mod. Phys.}\ }\textbf
  {\bibinfo {volume} {80}},\ \bibinfo {pages} {1083} (\bibinfo {year}
  {2008})}\BibitemShut {NoStop}%
\bibitem [{\citenamefont {Sarma}\ \emph {et~al.}(2015)\citenamefont {Sarma},
  \citenamefont {Freedman},\ and\ \citenamefont {Nayak}}]{Sarma:NQI15}%
  \BibitemOpen
  \bibfield  {author} {\bibinfo {author} {\bibfnamefont {S.~D.}\ \bibnamefont
  {Sarma}}, \bibinfo {author} {\bibfnamefont {M.}~\bibnamefont {Freedman}}, \
  and\ \bibinfo {author} {\bibfnamefont {C.}~\bibnamefont {Nayak}},\ }\href
  {http://dx.doi.org/10.1038/npjqi.2015.1} {\bibfield  {journal} {\bibinfo
  {journal} {Npj Quantum Information}\ }\textbf {\bibinfo {volume} {1}},\
  \bibinfo {pages} {15001 EP } (\bibinfo {year} {2015})}\BibitemShut {NoStop}%
\bibitem [{\citenamefont {Fu}\ and\ \citenamefont {Kane}(2008)}]{Fu:PRL08}%
  \BibitemOpen
  \bibfield  {author} {\bibinfo {author} {\bibfnamefont {L.}~\bibnamefont
  {Fu}}\ and\ \bibinfo {author} {\bibfnamefont {C.~L.}\ \bibnamefont {Kane}},\
  }\href {\doibase 10.1103/PhysRevLett.100.096407} {\bibfield  {journal}
  {\bibinfo  {journal} {Phys. Rev. Lett.}\ }\textbf {\bibinfo {volume} {100}},\
  \bibinfo {pages} {096407} (\bibinfo {year} {2008})}\BibitemShut {NoStop}%
\bibitem [{\citenamefont {Lutchyn}\ \emph {et~al.}(2010)\citenamefont
  {Lutchyn}, \citenamefont {Sau},\ and\ \citenamefont
  {Das~Sarma}}]{Lutchyn:PRL10}%
  \BibitemOpen
  \bibfield  {author} {\bibinfo {author} {\bibfnamefont {R.~M.}\ \bibnamefont
  {Lutchyn}}, \bibinfo {author} {\bibfnamefont {J.~D.}\ \bibnamefont {Sau}}, \
  and\ \bibinfo {author} {\bibfnamefont {S.}~\bibnamefont {Das~Sarma}},\ }\href
  {\doibase 10.1103/PhysRevLett.105.077001} {\bibfield  {journal} {\bibinfo
  {journal} {Phys. Rev. Lett.}\ }\textbf {\bibinfo {volume} {105}},\ \bibinfo
  {pages} {077001} (\bibinfo {year} {2010})}\BibitemShut {NoStop}%
\bibitem [{\citenamefont {Oreg}\ \emph {et~al.}(2010)\citenamefont {Oreg},
  \citenamefont {Refael},\ and\ \citenamefont {von Oppen}}]{Oreg:PRL10}%
  \BibitemOpen
  \bibfield  {author} {\bibinfo {author} {\bibfnamefont {Y.}~\bibnamefont
  {Oreg}}, \bibinfo {author} {\bibfnamefont {G.}~\bibnamefont {Refael}}, \ and\
  \bibinfo {author} {\bibfnamefont {F.}~\bibnamefont {von Oppen}},\ }\href
  {\doibase 10.1103/PhysRevLett.105.177002} {\bibfield  {journal} {\bibinfo
  {journal} {Phys. Rev. Lett.}\ }\textbf {\bibinfo {volume} {105}},\ \bibinfo
  {pages} {177002} (\bibinfo {year} {2010})}\BibitemShut {NoStop}%
\bibitem [{\citenamefont {Mourik}\ \emph {et~al.}(2012)\citenamefont {Mourik},
  \citenamefont {Zuo}, \citenamefont {Frolov}, \citenamefont {Plissard},
  \citenamefont {Bakkers},\ and\ \citenamefont {Kouwenhoven}}]{Mourik:S12}%
  \BibitemOpen
  \bibfield  {author} {\bibinfo {author} {\bibfnamefont {V.}~\bibnamefont
  {Mourik}}, \bibinfo {author} {\bibfnamefont {K.}~\bibnamefont {Zuo}},
  \bibinfo {author} {\bibfnamefont {S.~M.}\ \bibnamefont {Frolov}}, \bibinfo
  {author} {\bibfnamefont {S.~R.}\ \bibnamefont {Plissard}}, \bibinfo {author}
  {\bibfnamefont {E.~P. A.~M.}\ \bibnamefont {Bakkers}}, \ and\ \bibinfo
  {author} {\bibfnamefont {L.~P.}\ \bibnamefont {Kouwenhoven}},\ }\href
  {\doibase 10.1126/science.1222360} {\bibfield  {journal} {\bibinfo  {journal}
  {Science}\ }\textbf {\bibinfo {volume} {336}},\ \bibinfo {pages} {1003}
  (\bibinfo {year} {2012})}\BibitemShut {NoStop}%
\bibitem [{\citenamefont {Deng}\ \emph {et~al.}(2012)\citenamefont {Deng},
  \citenamefont {Yu}, \citenamefont {Huang}, \citenamefont {Larsson},
  \citenamefont {Caroff},\ and\ \citenamefont {Xu}}]{Deng:NL12}%
  \BibitemOpen
  \bibfield  {author} {\bibinfo {author} {\bibfnamefont {M.~T.}\ \bibnamefont
  {Deng}}, \bibinfo {author} {\bibfnamefont {C.~L.}\ \bibnamefont {Yu}},
  \bibinfo {author} {\bibfnamefont {G.~Y.}\ \bibnamefont {Huang}}, \bibinfo
  {author} {\bibfnamefont {M.}~\bibnamefont {Larsson}}, \bibinfo {author}
  {\bibfnamefont {P.}~\bibnamefont {Caroff}}, \ and\ \bibinfo {author}
  {\bibfnamefont {H.~Q.}\ \bibnamefont {Xu}},\ }\href {\doibase
  10.1021/nl303758w} {\bibfield  {journal} {\bibinfo  {journal} {Nano Lett.}\
  }\textbf {\bibinfo {volume} {12}},\ \bibinfo {pages} {6414} (\bibinfo {year}
  {2012})}\BibitemShut {NoStop}%
\bibitem [{\citenamefont {Das}\ \emph {et~al.}(2012)\citenamefont {Das},
  \citenamefont {Ronen}, \citenamefont {Most}, \citenamefont {Oreg},
  \citenamefont {Heiblum},\ and\ \citenamefont {Shtrikman}}]{Das:NP12}%
  \BibitemOpen
  \bibfield  {author} {\bibinfo {author} {\bibfnamefont {A.}~\bibnamefont
  {Das}}, \bibinfo {author} {\bibfnamefont {Y.}~\bibnamefont {Ronen}}, \bibinfo
  {author} {\bibfnamefont {Y.}~\bibnamefont {Most}}, \bibinfo {author}
  {\bibfnamefont {Y.}~\bibnamefont {Oreg}}, \bibinfo {author} {\bibfnamefont
  {M.}~\bibnamefont {Heiblum}}, \ and\ \bibinfo {author} {\bibfnamefont
  {H.}~\bibnamefont {Shtrikman}},\ }\href {http://dx.doi.org/10.1038/nphys2479}
  {\bibfield  {journal} {\bibinfo  {journal} {Nat. Phys.}\ }\textbf {\bibinfo
  {volume} {8}},\ \bibinfo {pages} {887} (\bibinfo {year} {2012})}\BibitemShut
  {NoStop}%
\bibitem [{\citenamefont {Churchill}\ \emph {et~al.}(2013)\citenamefont
  {Churchill}, \citenamefont {Fatemi}, \citenamefont {Grove-Rasmussen},
  \citenamefont {Deng}, \citenamefont {Caroff}, \citenamefont {Xu},\ and\
  \citenamefont {Marcus}}]{Churchill:PRB13}%
  \BibitemOpen
  \bibfield  {author} {\bibinfo {author} {\bibfnamefont {H.~O.~H.}\
  \bibnamefont {Churchill}}, \bibinfo {author} {\bibfnamefont {V.}~\bibnamefont
  {Fatemi}}, \bibinfo {author} {\bibfnamefont {K.}~\bibnamefont
  {Grove-Rasmussen}}, \bibinfo {author} {\bibfnamefont {M.~T.}\ \bibnamefont
  {Deng}}, \bibinfo {author} {\bibfnamefont {P.}~\bibnamefont {Caroff}},
  \bibinfo {author} {\bibfnamefont {H.~Q.}\ \bibnamefont {Xu}}, \ and\ \bibinfo
  {author} {\bibfnamefont {C.~M.}\ \bibnamefont {Marcus}},\ }\href {\doibase
  10.1103/PhysRevB.87.241401} {\bibfield  {journal} {\bibinfo  {journal} {Phys.
  Rev. B}\ }\textbf {\bibinfo {volume} {87}},\ \bibinfo {pages} {241401}
  (\bibinfo {year} {2013})}\BibitemShut {NoStop}%
\bibitem [{\citenamefont {Lee}\ \emph {et~al.}(2014)\citenamefont {Lee},
  \citenamefont {Jiang}, \citenamefont {Houzet}, \citenamefont {Aguado},
  \citenamefont {Lieber},\ and\ \citenamefont {De~Franceschi}}]{Lee:NN14}%
  \BibitemOpen
  \bibfield  {author} {\bibinfo {author} {\bibfnamefont {E.~J.~H.}\
  \bibnamefont {Lee}}, \bibinfo {author} {\bibfnamefont {X.}~\bibnamefont
  {Jiang}}, \bibinfo {author} {\bibfnamefont {M.}~\bibnamefont {Houzet}},
  \bibinfo {author} {\bibfnamefont {R.}~\bibnamefont {Aguado}}, \bibinfo
  {author} {\bibfnamefont {C.~M.}\ \bibnamefont {Lieber}}, \ and\ \bibinfo
  {author} {\bibfnamefont {S.}~\bibnamefont {De~Franceschi}},\ }\href
  {http://dx.doi.org/10.1038/nnano.2013.267} {\bibfield  {journal} {\bibinfo
  {journal} {Nat Nano}\ }\textbf {\bibinfo {volume} {9}},\ \bibinfo {pages}
  {79} (\bibinfo {year} {2014})}\BibitemShut {NoStop}%
\bibitem [{\citenamefont {Zhang}\ \emph {et~al.}(2016)\citenamefont {Zhang},
  \citenamefont {G{\"u}l}, \citenamefont {Conesa-Boj}, \citenamefont {Zuo},
  \citenamefont {Mourik}, \citenamefont {de~Vries}, \citenamefont {van Veen},
  \citenamefont {van Woerkom}, \citenamefont {Nowak}, \citenamefont {Wimmer},
  \citenamefont {Car}, \citenamefont {Plissard}, \citenamefont {Bakkers},
  \citenamefont {Quintero-P{\'e}rez}, \citenamefont {Goswami}, \citenamefont
  {Watanabe}, \citenamefont {Taniguchi},\ and\ \citenamefont
  {Kouwenhoven}}]{Zhang:16}%
  \BibitemOpen
  \bibfield  {author} {\bibinfo {author} {\bibfnamefont {H.}~\bibnamefont
  {Zhang}}, \bibinfo {author} {\bibfnamefont {{\"O}.}~\bibnamefont {G{\"u}l}},
  \bibinfo {author} {\bibfnamefont {S.}~\bibnamefont {Conesa-Boj}}, \bibinfo
  {author} {\bibfnamefont {K.}~\bibnamefont {Zuo}}, \bibinfo {author}
  {\bibfnamefont {V.}~\bibnamefont {Mourik}}, \bibinfo {author} {\bibfnamefont
  {F.~K.}\ \bibnamefont {de~Vries}}, \bibinfo {author} {\bibfnamefont
  {J.}~\bibnamefont {van Veen}}, \bibinfo {author} {\bibfnamefont {D.~J.}\
  \bibnamefont {van Woerkom}}, \bibinfo {author} {\bibfnamefont {M.~P.}\
  \bibnamefont {Nowak}}, \bibinfo {author} {\bibfnamefont {M.}~\bibnamefont
  {Wimmer}}, \bibinfo {author} {\bibfnamefont {D.}~\bibnamefont {Car}},
  \bibinfo {author} {\bibfnamefont {S.}~\bibnamefont {Plissard}}, \bibinfo
  {author} {\bibfnamefont {E.~P. A.~M.}\ \bibnamefont {Bakkers}}, \bibinfo
  {author} {\bibfnamefont {M.}~\bibnamefont {Quintero-P{\'e}rez}}, \bibinfo
  {author} {\bibfnamefont {S.}~\bibnamefont {Goswami}}, \bibinfo {author}
  {\bibfnamefont {K.}~\bibnamefont {Watanabe}}, \bibinfo {author}
  {\bibfnamefont {T.}~\bibnamefont {Taniguchi}}, \ and\ \bibinfo {author}
  {\bibfnamefont {L.~P.}\ \bibnamefont {Kouwenhoven}},\ }\href
  {http://arxiv.org/abs/1603.04069} {\  (\bibinfo {year} {2016})},\ \Eprint
  {http://arxiv.org/abs/arXiv:1603.04069} {arXiv:1603.04069} \BibitemShut
  {NoStop}%
\bibitem [{\citenamefont {Deng}\ \emph {et~al.}(2016)\citenamefont {Deng},
  \citenamefont {Vaitiekenas}, \citenamefont {Hansen}, \citenamefont {Danon},
  \citenamefont {Leijnse}, \citenamefont {Flensberg}, \citenamefont
  {Nyg{\aa}rd}, \citenamefont {Krogstrup},\ and\ \citenamefont
  {Marcus}}]{Deng:S16}%
  \BibitemOpen
  \bibfield  {author} {\bibinfo {author} {\bibfnamefont {M.~T.}\ \bibnamefont
  {Deng}}, \bibinfo {author} {\bibfnamefont {S.}~\bibnamefont {Vaitiekenas}},
  \bibinfo {author} {\bibfnamefont {E.~B.}\ \bibnamefont {Hansen}}, \bibinfo
  {author} {\bibfnamefont {J.}~\bibnamefont {Danon}}, \bibinfo {author}
  {\bibfnamefont {M.}~\bibnamefont {Leijnse}}, \bibinfo {author} {\bibfnamefont
  {K.}~\bibnamefont {Flensberg}}, \bibinfo {author} {\bibfnamefont
  {J.}~\bibnamefont {Nyg{\aa}rd}}, \bibinfo {author} {\bibfnamefont
  {P.}~\bibnamefont {Krogstrup}}, \ and\ \bibinfo {author} {\bibfnamefont
  {C.~M.}\ \bibnamefont {Marcus}},\ }\href {\doibase 10.1126/science.aaf3961}
  {\bibfield  {journal} {\bibinfo  {journal} {Science}\ }\textbf {\bibinfo
  {volume} {354}},\ \bibinfo {pages} {1557} (\bibinfo {year}
  {2016})}\BibitemShut {NoStop}%
\bibitem [{\citenamefont {Liu}\ \emph {et~al.}(2012)\citenamefont {Liu},
  \citenamefont {Potter}, \citenamefont {Law},\ and\ \citenamefont
  {Lee}}]{Liu:PRL12}%
  \BibitemOpen
  \bibfield  {author} {\bibinfo {author} {\bibfnamefont {J.}~\bibnamefont
  {Liu}}, \bibinfo {author} {\bibfnamefont {A.~C.}\ \bibnamefont {Potter}},
  \bibinfo {author} {\bibfnamefont {K.~T.}\ \bibnamefont {Law}}, \ and\
  \bibinfo {author} {\bibfnamefont {P.~A.}\ \bibnamefont {Lee}},\ }\href
  {\doibase 10.1103/PhysRevLett.109.267002} {\bibfield  {journal} {\bibinfo
  {journal} {Phys. Rev. Lett.}\ }\textbf {\bibinfo {volume} {109}},\ \bibinfo
  {pages} {267002} (\bibinfo {year} {2012})}\BibitemShut {NoStop}%
\bibitem [{\citenamefont {Pikulin}\ \emph {et~al.}(2012)\citenamefont
  {Pikulin}, \citenamefont {Dahlhaus}, \citenamefont {Wimmer}, \citenamefont
  {Schomerus},\ and\ \citenamefont {Beenakker}}]{Pikulin:NJP12}%
  \BibitemOpen
  \bibfield  {author} {\bibinfo {author} {\bibfnamefont {D.~I.}\ \bibnamefont
  {Pikulin}}, \bibinfo {author} {\bibfnamefont {J.~P.}\ \bibnamefont
  {Dahlhaus}}, \bibinfo {author} {\bibfnamefont {M.}~\bibnamefont {Wimmer}},
  \bibinfo {author} {\bibfnamefont {H.}~\bibnamefont {Schomerus}}, \ and\
  \bibinfo {author} {\bibfnamefont {C.~W.~J.}\ \bibnamefont {Beenakker}},\
  }\href {http://stacks.iop.org/1367-2630/14/i=12/a=125011} {\bibfield
  {journal} {\bibinfo  {journal} {New J. Phys.}\ }\textbf {\bibinfo {volume}
  {14}},\ \bibinfo {pages} {125011} (\bibinfo {year} {2012})}\BibitemShut
  {NoStop}%
\bibitem [{\citenamefont {Lee}\ \emph {et~al.}(2012)\citenamefont {Lee},
  \citenamefont {Jiang}, \citenamefont {Aguado}, \citenamefont {Katsaros},
  \citenamefont {Lieber},\ and\ \citenamefont {De~Franceschi}}]{Lee:PRL12}%
  \BibitemOpen
  \bibfield  {author} {\bibinfo {author} {\bibfnamefont {E.~J.~H.}\
  \bibnamefont {Lee}}, \bibinfo {author} {\bibfnamefont {X.}~\bibnamefont
  {Jiang}}, \bibinfo {author} {\bibfnamefont {R.}~\bibnamefont {Aguado}},
  \bibinfo {author} {\bibfnamefont {G.}~\bibnamefont {Katsaros}}, \bibinfo
  {author} {\bibfnamefont {C.~M.}\ \bibnamefont {Lieber}}, \ and\ \bibinfo
  {author} {\bibfnamefont {S.}~\bibnamefont {De~Franceschi}},\ }\href {\doibase
  10.1103/PhysRevLett.109.186802} {\bibfield  {journal} {\bibinfo  {journal}
  {Phys. Rev. Lett.}\ }\textbf {\bibinfo {volume} {109}},\ \bibinfo {pages}
  {186802} (\bibinfo {year} {2012})}\BibitemShut {NoStop}%
\bibitem [{\citenamefont {Zitko}\ \emph {et~al.}(2015)\citenamefont {Zitko},
  \citenamefont {Lim}, \citenamefont {L\'opez},\ and\ \citenamefont
  {Aguado}}]{Zitko:PRB15}%
  \BibitemOpen
  \bibfield  {author} {\bibinfo {author} {\bibfnamefont {R.}~\bibnamefont
  {Zitko}}, \bibinfo {author} {\bibfnamefont {J.~S.}\ \bibnamefont {Lim}},
  \bibinfo {author} {\bibfnamefont {R.}~\bibnamefont {L\'opez}}, \ and\
  \bibinfo {author} {\bibfnamefont {R.}~\bibnamefont {Aguado}},\ }\href
  {\doibase 10.1103/PhysRevB.91.045441} {\bibfield  {journal} {\bibinfo
  {journal} {Phys. Rev. B}\ }\textbf {\bibinfo {volume} {91}},\ \bibinfo
  {pages} {045441} (\bibinfo {year} {2015})}\BibitemShut {NoStop}%
\bibitem [{\citenamefont {Nilsson}\ \emph {et~al.}(2008)\citenamefont
  {Nilsson}, \citenamefont {Akhmerov},\ and\ \citenamefont
  {Beenakker}}]{Nilsson:PRL08}%
  \BibitemOpen
  \bibfield  {author} {\bibinfo {author} {\bibfnamefont {J.}~\bibnamefont
  {Nilsson}}, \bibinfo {author} {\bibfnamefont {A.~R.}\ \bibnamefont
  {Akhmerov}}, \ and\ \bibinfo {author} {\bibfnamefont {C.~W.~J.}\ \bibnamefont
  {Beenakker}},\ }\href {\doibase 10.1103/PhysRevLett.101.120403} {\bibfield
  {journal} {\bibinfo  {journal} {Phys. Rev. Lett.}\ }\textbf {\bibinfo
  {volume} {101}},\ \bibinfo {pages} {120403} (\bibinfo {year}
  {2008})}\BibitemShut {NoStop}%
\bibitem [{\citenamefont {Tewari}\ \emph {et~al.}(2008)\citenamefont {Tewari},
  \citenamefont {Zhang}, \citenamefont {Das~Sarma}, \citenamefont {Nayak},\
  and\ \citenamefont {Lee}}]{Tewari:PRL08}%
  \BibitemOpen
  \bibfield  {author} {\bibinfo {author} {\bibfnamefont {S.}~\bibnamefont
  {Tewari}}, \bibinfo {author} {\bibfnamefont {C.}~\bibnamefont {Zhang}},
  \bibinfo {author} {\bibfnamefont {S.}~\bibnamefont {Das~Sarma}}, \bibinfo
  {author} {\bibfnamefont {C.}~\bibnamefont {Nayak}}, \ and\ \bibinfo {author}
  {\bibfnamefont {D.-H.}\ \bibnamefont {Lee}},\ }\href {\doibase
  10.1103/PhysRevLett.100.027001} {\bibfield  {journal} {\bibinfo  {journal}
  {Phys. Rev. Lett.}\ }\textbf {\bibinfo {volume} {100}},\ \bibinfo {pages}
  {027001} (\bibinfo {year} {2008})}\BibitemShut {NoStop}%
\bibitem [{\citenamefont {Fu}(2010)}]{Fu:PRL10}%
  \BibitemOpen
  \bibfield  {author} {\bibinfo {author} {\bibfnamefont {L.}~\bibnamefont
  {Fu}},\ }\href {\doibase 10.1103/PhysRevLett.104.056402} {\bibfield
  {journal} {\bibinfo  {journal} {Phys. Rev. Lett.}\ }\textbf {\bibinfo
  {volume} {104}},\ \bibinfo {pages} {056402} (\bibinfo {year}
  {2010})}\BibitemShut {NoStop}%
\bibitem [{\citenamefont {Alicea}\ \emph {et~al.}(2011)\citenamefont {Alicea},
  \citenamefont {Oreg}, \citenamefont {Refael}, \citenamefont {von Oppen},\
  and\ \citenamefont {Fisher}}]{Alicea:NP11}%
  \BibitemOpen
  \bibfield  {author} {\bibinfo {author} {\bibfnamefont {J.}~\bibnamefont
  {Alicea}}, \bibinfo {author} {\bibfnamefont {Y.}~\bibnamefont {Oreg}},
  \bibinfo {author} {\bibfnamefont {G.}~\bibnamefont {Refael}}, \bibinfo
  {author} {\bibfnamefont {F.}~\bibnamefont {von Oppen}}, \ and\ \bibinfo
  {author} {\bibfnamefont {M.~P.~A.}\ \bibnamefont {Fisher}},\ }\href
  {http://dx.doi.org/10.1038/nphys1915} {\bibfield  {journal} {\bibinfo
  {journal} {Nat. Phys.}\ }\textbf {\bibinfo {volume} {7}},\ \bibinfo {pages}
  {412} (\bibinfo {year} {2011})}\BibitemShut {NoStop}%
\bibitem [{\citenamefont {Sau}\ \emph {et~al.}(2011)\citenamefont {Sau},
  \citenamefont {Clarke},\ and\ \citenamefont {Tewari}}]{Sau:PRB11}%
  \BibitemOpen
  \bibfield  {author} {\bibinfo {author} {\bibfnamefont {J.~D.}\ \bibnamefont
  {Sau}}, \bibinfo {author} {\bibfnamefont {D.~J.}\ \bibnamefont {Clarke}}, \
  and\ \bibinfo {author} {\bibfnamefont {S.}~\bibnamefont {Tewari}},\ }\href
  {\doibase 10.1103/PhysRevB.84.094505} {\bibfield  {journal} {\bibinfo
  {journal} {Phys. Rev. B}\ }\textbf {\bibinfo {volume} {84}},\ \bibinfo
  {pages} {094505} (\bibinfo {year} {2011})}\BibitemShut {NoStop}%
\bibitem [{\citenamefont {Halperin}\ \emph {et~al.}(2012)\citenamefont
  {Halperin}, \citenamefont {Oreg}, \citenamefont {Stern}, \citenamefont
  {Refael}, \citenamefont {Alicea},\ and\ \citenamefont {von
  Oppen}}]{Halperin:PRB12}%
  \BibitemOpen
  \bibfield  {author} {\bibinfo {author} {\bibfnamefont {B.~I.}\ \bibnamefont
  {Halperin}}, \bibinfo {author} {\bibfnamefont {Y.}~\bibnamefont {Oreg}},
  \bibinfo {author} {\bibfnamefont {A.}~\bibnamefont {Stern}}, \bibinfo
  {author} {\bibfnamefont {G.}~\bibnamefont {Refael}}, \bibinfo {author}
  {\bibfnamefont {J.}~\bibnamefont {Alicea}}, \ and\ \bibinfo {author}
  {\bibfnamefont {F.}~\bibnamefont {von Oppen}},\ }\href {\doibase
  10.1103/PhysRevB.85.144501} {\bibfield  {journal} {\bibinfo  {journal} {Phys.
  Rev. B}\ }\textbf {\bibinfo {volume} {85}},\ \bibinfo {pages} {144501}
  (\bibinfo {year} {2012})}\BibitemShut {NoStop}%
\bibitem [{\citenamefont {van Heck}\ \emph {et~al.}(2012)\citenamefont {van
  Heck}, \citenamefont {Akhmerov}, \citenamefont {Hassler}, \citenamefont
  {Burrello},\ and\ \citenamefont {Beenakker}}]{Heck:NJP12}%
  \BibitemOpen
  \bibfield  {author} {\bibinfo {author} {\bibfnamefont {B.}~\bibnamefont {van
  Heck}}, \bibinfo {author} {\bibfnamefont {A.~R.}\ \bibnamefont {Akhmerov}},
  \bibinfo {author} {\bibfnamefont {F.}~\bibnamefont {Hassler}}, \bibinfo
  {author} {\bibfnamefont {M.}~\bibnamefont {Burrello}}, \ and\ \bibinfo
  {author} {\bibfnamefont {C.~W.~J.}\ \bibnamefont {Beenakker}},\ }\href
  {http://stacks.iop.org/1367-2630/14/i=3/a=035019} {\bibfield  {journal}
  {\bibinfo  {journal} {New J. Phys.}\ }\textbf {\bibinfo {volume} {14}},\
  \bibinfo {pages} {035019} (\bibinfo {year} {2012})}\BibitemShut {NoStop}%
\bibitem [{\citenamefont {Hyart}\ \emph {et~al.}(2013)\citenamefont {Hyart},
  \citenamefont {van Heck}, \citenamefont {Fulga}, \citenamefont {Burrello},
  \citenamefont {Akhmerov},\ and\ \citenamefont {Beenakker}}]{Hyart:PRB13}%
  \BibitemOpen
  \bibfield  {author} {\bibinfo {author} {\bibfnamefont {T.}~\bibnamefont
  {Hyart}}, \bibinfo {author} {\bibfnamefont {B.}~\bibnamefont {van Heck}},
  \bibinfo {author} {\bibfnamefont {I.~C.}\ \bibnamefont {Fulga}}, \bibinfo
  {author} {\bibfnamefont {M.}~\bibnamefont {Burrello}}, \bibinfo {author}
  {\bibfnamefont {A.~R.}\ \bibnamefont {Akhmerov}}, \ and\ \bibinfo {author}
  {\bibfnamefont {C.~W.~J.}\ \bibnamefont {Beenakker}},\ }\href {\doibase
  10.1103/PhysRevB.88.035121} {\bibfield  {journal} {\bibinfo  {journal} {Phys.
  Rev. B}\ }\textbf {\bibinfo {volume} {88}},\ \bibinfo {pages} {035121}
  (\bibinfo {year} {2013})}\BibitemShut {NoStop}%
\bibitem [{\citenamefont {Liu}\ \emph {et~al.}(2013)\citenamefont {Liu},
  \citenamefont {Zhang},\ and\ \citenamefont {Law}}]{Liu:PRB13}%
  \BibitemOpen
  \bibfield  {author} {\bibinfo {author} {\bibfnamefont {J.}~\bibnamefont
  {Liu}}, \bibinfo {author} {\bibfnamefont {F.-C.}\ \bibnamefont {Zhang}}, \
  and\ \bibinfo {author} {\bibfnamefont {K.~T.}\ \bibnamefont {Law}},\ }\href
  {\doibase 10.1103/PhysRevB.88.064509} {\bibfield  {journal} {\bibinfo
  {journal} {Phys. Rev. B}\ }\textbf {\bibinfo {volume} {88}},\ \bibinfo
  {pages} {064509} (\bibinfo {year} {2013})}\BibitemShut {NoStop}%
\bibitem [{\citenamefont {Zocher}\ and\ \citenamefont
  {Rosenow}(2013)}]{Zocher:PRL13}%
  \BibitemOpen
  \bibfield  {author} {\bibinfo {author} {\bibfnamefont {B.}~\bibnamefont
  {Zocher}}\ and\ \bibinfo {author} {\bibfnamefont {B.}~\bibnamefont
  {Rosenow}},\ }\href {\doibase 10.1103/PhysRevLett.111.036802} {\bibfield
  {journal} {\bibinfo  {journal} {Phys. Rev. Lett.}\ }\textbf {\bibinfo
  {volume} {111}},\ \bibinfo {pages} {036802} (\bibinfo {year}
  {2013})}\BibitemShut {NoStop}%
\bibitem [{\citenamefont {Fregoso}\ \emph {et~al.}(2013)\citenamefont
  {Fregoso}, \citenamefont {Lobos},\ and\ \citenamefont
  {Das~Sarma}}]{Fregoso:PRB13}%
  \BibitemOpen
  \bibfield  {author} {\bibinfo {author} {\bibfnamefont {B.~M.}\ \bibnamefont
  {Fregoso}}, \bibinfo {author} {\bibfnamefont {A.~M.}\ \bibnamefont {Lobos}},
  \ and\ \bibinfo {author} {\bibfnamefont {S.}~\bibnamefont {Das~Sarma}},\
  }\href {\doibase 10.1103/PhysRevB.88.180507} {\bibfield  {journal} {\bibinfo
  {journal} {Phys. Rev. B}\ }\textbf {\bibinfo {volume} {88}},\ \bibinfo
  {pages} {180507} (\bibinfo {year} {2013})}\BibitemShut {NoStop}%
\bibitem [{\citenamefont {Li}\ \emph {et~al.}(2014)\citenamefont {Li},
  \citenamefont {Yu}, \citenamefont {Lin},\ and\ \citenamefont
  {You}}]{Li:SR14}%
  \BibitemOpen
  \bibfield  {author} {\bibinfo {author} {\bibfnamefont {J.}~\bibnamefont
  {Li}}, \bibinfo {author} {\bibfnamefont {T.}~\bibnamefont {Yu}}, \bibinfo
  {author} {\bibfnamefont {H.-Q.}\ \bibnamefont {Lin}}, \ and\ \bibinfo
  {author} {\bibfnamefont {J.~Q.}\ \bibnamefont {You}},\ }\href
  {http://dx.doi.org/10.1038/srep04930} {\bibfield  {journal} {\bibinfo
  {journal} {Sci. Rep.}\ }\textbf {\bibinfo {volume} {4}} (\bibinfo {year}
  {2014})}\BibitemShut {NoStop}%
\bibitem [{\citenamefont {Lobos}\ and\ \citenamefont
  {Sarma}(2015)}]{Lobos:NJP15}%
  \BibitemOpen
  \bibfield  {author} {\bibinfo {author} {\bibfnamefont {A.~M.}\ \bibnamefont
  {Lobos}}\ and\ \bibinfo {author} {\bibfnamefont {S.~D.}\ \bibnamefont
  {Sarma}},\ }\href {http://stacks.iop.org/1367-2630/17/i=6/a=065010}
  {\bibfield  {journal} {\bibinfo  {journal} {New J. Phys.}\ }\textbf {\bibinfo
  {volume} {17}},\ \bibinfo {pages} {065010} (\bibinfo {year}
  {2015})}\BibitemShut {NoStop}%
\bibitem [{\citenamefont {Haim}\ \emph {et~al.}(2015)\citenamefont {Haim},
  \citenamefont {Berg}, \citenamefont {von Oppen},\ and\ \citenamefont
  {Oreg}}]{Haim:PRL15}%
  \BibitemOpen
  \bibfield  {author} {\bibinfo {author} {\bibfnamefont {A.}~\bibnamefont
  {Haim}}, \bibinfo {author} {\bibfnamefont {E.}~\bibnamefont {Berg}}, \bibinfo
  {author} {\bibfnamefont {F.}~\bibnamefont {von Oppen}}, \ and\ \bibinfo
  {author} {\bibfnamefont {Y.}~\bibnamefont {Oreg}},\ }\href {\doibase
  10.1103/PhysRevLett.114.166406} {\bibfield  {journal} {\bibinfo  {journal}
  {Phys. Rev. Lett.}\ }\textbf {\bibinfo {volume} {114}},\ \bibinfo {pages}
  {166406} (\bibinfo {year} {2015})}\BibitemShut {NoStop}%
\bibitem [{\citenamefont {Aasen}\ \emph {et~al.}(2016)\citenamefont {Aasen},
  \citenamefont {Hell}, \citenamefont {Mishmash}, \citenamefont {Higginbotham},
  \citenamefont {Danon}, \citenamefont {Leijnse}, \citenamefont {Jespersen},
  \citenamefont {Folk}, \citenamefont {Marcus}, \citenamefont {Flensberg},\
  and\ \citenamefont {Alicea}}]{Aasen:PRX16}%
  \BibitemOpen
  \bibfield  {author} {\bibinfo {author} {\bibfnamefont {D.}~\bibnamefont
  {Aasen}}, \bibinfo {author} {\bibfnamefont {M.}~\bibnamefont {Hell}},
  \bibinfo {author} {\bibfnamefont {R.~V.}\ \bibnamefont {Mishmash}}, \bibinfo
  {author} {\bibfnamefont {A.}~\bibnamefont {Higginbotham}}, \bibinfo {author}
  {\bibfnamefont {J.}~\bibnamefont {Danon}}, \bibinfo {author} {\bibfnamefont
  {M.}~\bibnamefont {Leijnse}}, \bibinfo {author} {\bibfnamefont {T.~S.}\
  \bibnamefont {Jespersen}}, \bibinfo {author} {\bibfnamefont {J.~A.}\
  \bibnamefont {Folk}}, \bibinfo {author} {\bibfnamefont {C.~M.}\ \bibnamefont
  {Marcus}}, \bibinfo {author} {\bibfnamefont {K.}~\bibnamefont {Flensberg}}, \
  and\ \bibinfo {author} {\bibfnamefont {J.}~\bibnamefont {Alicea}},\ }\href
  {\doibase 10.1103/PhysRevX.6.031016} {\bibfield  {journal} {\bibinfo
  {journal} {Phys. Rev. X}\ }\textbf {\bibinfo {volume} {6}},\ \bibinfo {pages}
  {031016} (\bibinfo {year} {2016})}\BibitemShut {NoStop}%
\bibitem [{\citenamefont {Chiu}\ \emph {et~al.}(2015)\citenamefont {Chiu},
  \citenamefont {Vazifeh},\ and\ \citenamefont {Franz}}]{Chiu:EEL15}%
  \BibitemOpen
  \bibfield  {author} {\bibinfo {author} {\bibfnamefont {C.-K.}\ \bibnamefont
  {Chiu}}, \bibinfo {author} {\bibfnamefont {M.~M.}\ \bibnamefont {Vazifeh}}, \
  and\ \bibinfo {author} {\bibfnamefont {M.}~\bibnamefont {Franz}},\ }\href
  {http://stacks.iop.org/0295-5075/110/i=1/a=10001} {\bibfield  {journal}
  {\bibinfo  {journal} {EPL (Europhysics Letters)}\ }\textbf {\bibinfo {volume}
  {110}},\ \bibinfo {pages} {10001} (\bibinfo {year} {2015})}\BibitemShut
  {NoStop}%
\bibitem [{\citenamefont {Vijay}\ and\ \citenamefont {Fu}(2016)}]{Vijay:PRB16}%
  \BibitemOpen
  \bibfield  {author} {\bibinfo {author} {\bibfnamefont {S.}~\bibnamefont
  {Vijay}}\ and\ \bibinfo {author} {\bibfnamefont {L.}~\bibnamefont {Fu}},\
  }\href {\doibase 10.1103/PhysRevB.94.235446} {\bibfield  {journal} {\bibinfo
  {journal} {Phys. Rev. B}\ }\textbf {\bibinfo {volume} {94}},\ \bibinfo
  {pages} {235446} (\bibinfo {year} {2016})}\BibitemShut {NoStop}%
\bibitem [{\citenamefont {Karzig}\ \emph {et~al.}(2016)\citenamefont {Karzig},
  \citenamefont {Knapp}, \citenamefont {Lutchyn}, \citenamefont {Bonderson},
  \citenamefont {Hastings}, \citenamefont {Nayak}, \citenamefont {Alicea},
  \citenamefont {Flensberg}, \citenamefont {Plugge}, \citenamefont {Oreg},
  \citenamefont {Marcus},\ and\ \citenamefont {Freedman}}]{Karzig:16}%
  \BibitemOpen
  \bibfield  {author} {\bibinfo {author} {\bibfnamefont {T.}~\bibnamefont
  {Karzig}}, \bibinfo {author} {\bibfnamefont {C.}~\bibnamefont {Knapp}},
  \bibinfo {author} {\bibfnamefont {R.}~\bibnamefont {Lutchyn}}, \bibinfo
  {author} {\bibfnamefont {P.}~\bibnamefont {Bonderson}}, \bibinfo {author}
  {\bibfnamefont {M.}~\bibnamefont {Hastings}}, \bibinfo {author}
  {\bibfnamefont {C.}~\bibnamefont {Nayak}}, \bibinfo {author} {\bibfnamefont
  {J.}~\bibnamefont {Alicea}}, \bibinfo {author} {\bibfnamefont
  {K.}~\bibnamefont {Flensberg}}, \bibinfo {author} {\bibfnamefont
  {S.}~\bibnamefont {Plugge}}, \bibinfo {author} {\bibfnamefont
  {Y.}~\bibnamefont {Oreg}}, \bibinfo {author} {\bibfnamefont {C.}~\bibnamefont
  {Marcus}}, \ and\ \bibinfo {author} {\bibfnamefont {M.~H.}\ \bibnamefont
  {Freedman}},\ }\href {https://arxiv.org/abs/1610.05289} {\  (\bibinfo {year}
  {2016})},\ \Eprint {http://arxiv.org/abs/arXiv:1610.05289} {arXiv:1610.05289}
  \BibitemShut {NoStop}%
\bibitem [{\citenamefont {Plugge}\ \emph {et~al.}(2017)\citenamefont {Plugge},
  \citenamefont {Rasmussen}, \citenamefont {Egger},\ and\ \citenamefont
  {Flensberg}}]{Plugge:NJP17}%
  \BibitemOpen
  \bibfield  {author} {\bibinfo {author} {\bibfnamefont {S.}~\bibnamefont
  {Plugge}}, \bibinfo {author} {\bibfnamefont {A.}~\bibnamefont {Rasmussen}},
  \bibinfo {author} {\bibfnamefont {R.}~\bibnamefont {Egger}}, \ and\ \bibinfo
  {author} {\bibfnamefont {K.}~\bibnamefont {Flensberg}},\ }\href {\doibase
  10.1088/1367-2630/aa54e1} {\bibfield  {journal} {\bibinfo  {journal} {New J.
  Phys.}\ }\textbf {\bibinfo {volume} {19}},\ \bibinfo {pages} {012001}
  (\bibinfo {year} {2017})}\BibitemShut {NoStop}%
\bibitem [{\citenamefont {Mishmash}\ \emph {et~al.}(2016)\citenamefont
  {Mishmash}, \citenamefont {Aasen}, \citenamefont {Higginbotham},\ and\
  \citenamefont {Alicea}}]{Mishmash:PRB16}%
  \BibitemOpen
  \bibfield  {author} {\bibinfo {author} {\bibfnamefont {R.~V.}\ \bibnamefont
  {Mishmash}}, \bibinfo {author} {\bibfnamefont {D.}~\bibnamefont {Aasen}},
  \bibinfo {author} {\bibfnamefont {A.~P.}\ \bibnamefont {Higginbotham}}, \
  and\ \bibinfo {author} {\bibfnamefont {J.}~\bibnamefont {Alicea}},\ }\href
  {\doibase 10.1103/PhysRevB.93.245404} {\bibfield  {journal} {\bibinfo
  {journal} {Phys. Rev. B}\ }\textbf {\bibinfo {volume} {93}},\ \bibinfo
  {pages} {245404} (\bibinfo {year} {2016})}\BibitemShut {NoStop}%
\bibitem [{Note1()}]{Note1}%
  \BibitemOpen
  \bibinfo {note} {It is well known that mean field solutions artificially
  break time reversal symmetry. However, we always focus on the large Zeeman
  regime, which is well described by a mean field approximation. Furthermore,
  we note that Kondo physics is only relevant in the regime $T_K/\Delta \gtrsim
  0.6$ (see e.g. E. J. H. Lee {\protect \it et al}, Phys. Rev. B {\protect \bf
  95}, 180502(R), 2017) which is far from the parameter regime we consider
  (with a doublet ground state for odd occupancy, i. e. at the center of the
  Coulomb Blockade diamond).}\BibitemShut {Stop}%
\bibitem [{\citenamefont {Leijnse}\ and\ \citenamefont
  {Flensberg}(2011)}]{Leijnse:PRB11}%
  \BibitemOpen
  \bibfield  {author} {\bibinfo {author} {\bibfnamefont {M.}~\bibnamefont
  {Leijnse}}\ and\ \bibinfo {author} {\bibfnamefont {K.}~\bibnamefont
  {Flensberg}},\ }\href {\doibase 10.1103/PhysRevB.84.140501} {\bibfield
  {journal} {\bibinfo  {journal} {Phys. Rev. B}\ }\textbf {\bibinfo {volume}
  {84}},\ \bibinfo {pages} {140501} (\bibinfo {year} {2011})}\BibitemShut
  {NoStop}%
\bibitem [{\citenamefont {Prada}\ \emph {et~al.}(2012)\citenamefont {Prada},
  \citenamefont {San-Jose},\ and\ \citenamefont {Aguado}}]{Prada:PRB12}%
  \BibitemOpen
  \bibfield  {author} {\bibinfo {author} {\bibfnamefont {E.}~\bibnamefont
  {Prada}}, \bibinfo {author} {\bibfnamefont {P.}~\bibnamefont {San-Jose}}, \
  and\ \bibinfo {author} {\bibfnamefont {R.}~\bibnamefont {Aguado}},\ }\href
  {\doibase 10.1103/PhysRevB.86.180503} {\bibfield  {journal} {\bibinfo
  {journal} {Phys. Rev. B}\ }\textbf {\bibinfo {volume} {86}},\ \bibinfo
  {pages} {180503(R)} (\bibinfo {year} {2012})}\BibitemShut {NoStop}%
\bibitem [{\citenamefont {Vernek}\ \emph {et~al.}(2014)\citenamefont {Vernek},
  \citenamefont {Penteado}, \citenamefont {Seridonio},\ and\ \citenamefont
  {Egues}}]{Vernek:PRB14}%
  \BibitemOpen
  \bibfield  {author} {\bibinfo {author} {\bibfnamefont {E.}~\bibnamefont
  {Vernek}}, \bibinfo {author} {\bibfnamefont {P.~H.}\ \bibnamefont
  {Penteado}}, \bibinfo {author} {\bibfnamefont {A.~C.}\ \bibnamefont
  {Seridonio}}, \ and\ \bibinfo {author} {\bibfnamefont {J.~C.}\ \bibnamefont
  {Egues}},\ }\href {\doibase 10.1103/PhysRevB.89.165314} {\bibfield  {journal}
  {\bibinfo  {journal} {Phys. Rev. B}\ }\textbf {\bibinfo {volume} {89}},\
  \bibinfo {pages} {165314} (\bibinfo {year} {2014})}\BibitemShut {NoStop}%
\bibitem [{\citenamefont {Ruiz-Tijerina}\ \emph {et~al.}(2015)\citenamefont
  {Ruiz-Tijerina}, \citenamefont {Vernek}, \citenamefont {Dias~da Silva},\ and\
  \citenamefont {Egues}}]{Ruiz-Tijerina:PRB15}%
  \BibitemOpen
  \bibfield  {author} {\bibinfo {author} {\bibfnamefont {D.~A.}\ \bibnamefont
  {Ruiz-Tijerina}}, \bibinfo {author} {\bibfnamefont {E.}~\bibnamefont
  {Vernek}}, \bibinfo {author} {\bibfnamefont {L.~G. G.~V.}\ \bibnamefont
  {Dias~da Silva}}, \ and\ \bibinfo {author} {\bibfnamefont {J.~C.}\
  \bibnamefont {Egues}},\ }\href {\doibase 10.1103/PhysRevB.91.115435}
  {\bibfield  {journal} {\bibinfo  {journal} {Phys. Rev. B}\ }\textbf {\bibinfo
  {volume} {91}},\ \bibinfo {pages} {115435} (\bibinfo {year}
  {2015})}\BibitemShut {NoStop}%
\bibitem [{\citenamefont {Cayao}\ \emph {et~al.}(2015)\citenamefont {Cayao},
  \citenamefont {Prada}, \citenamefont {San-Jose},\ and\ \citenamefont
  {Aguado}}]{Cayao:PRB15}%
  \BibitemOpen
  \bibfield  {author} {\bibinfo {author} {\bibfnamefont {J.}~\bibnamefont
  {Cayao}}, \bibinfo {author} {\bibfnamefont {E.}~\bibnamefont {Prada}},
  \bibinfo {author} {\bibfnamefont {P.}~\bibnamefont {San-Jose}}, \ and\
  \bibinfo {author} {\bibfnamefont {R.}~\bibnamefont {Aguado}},\ }\href
  {\doibase 10.1103/PhysRevB.91.024514} {\bibfield  {journal} {\bibinfo
  {journal} {Phys. Rev. B}\ }\textbf {\bibinfo {volume} {91}},\ \bibinfo
  {pages} {024514} (\bibinfo {year} {2015})}\BibitemShut {NoStop}%
\bibitem [{\citenamefont {Ricco}\ \emph {et~al.}(2016)\citenamefont {Ricco},
  \citenamefont {Dessotti}, \citenamefont {Ramalho}, \citenamefont {Figueira},\
  and\ \citenamefont {Seridonio}}]{Ricco:16}%
  \BibitemOpen
  \bibfield  {author} {\bibinfo {author} {\bibfnamefont {L.~S.}\ \bibnamefont
  {Ricco}}, \bibinfo {author} {\bibfnamefont {F.~A.}\ \bibnamefont {Dessotti}},
  \bibinfo {author} {\bibfnamefont {A.}~\bibnamefont {Ramalho}}, \bibinfo
  {author} {\bibfnamefont {M.~S.}\ \bibnamefont {Figueira}}, \ and\ \bibinfo
  {author} {\bibfnamefont {A.~C.}\ \bibnamefont {Seridonio}},\ }\href
  {https://arxiv.org/abs/1611.04347} {\  (\bibinfo {year} {2016})},\ \Eprint
  {http://arxiv.org/abs/arXiv:1611.04347} {arXiv:1611.04347} \BibitemShut
  {NoStop}%
\bibitem [{\citenamefont {Sau}\ and\ \citenamefont
  {Setiawan}(2017)}]{Sau:PRB17}%
  \BibitemOpen
  \bibfield  {author} {\bibinfo {author} {\bibfnamefont {J.~D.}\ \bibnamefont
  {Sau}}\ and\ \bibinfo {author} {\bibfnamefont {F.}~\bibnamefont {Setiawan}},\
  }\href {\doibase 10.1103/PhysRevB.95.060501} {\bibfield  {journal} {\bibinfo
  {journal} {Phys. Rev. B}\ }\textbf {\bibinfo {volume} {95}},\ \bibinfo
  {pages} {060501} (\bibinfo {year} {2017})}\BibitemShut {NoStop}%
\bibitem [{\citenamefont {Klinovaja}\ and\ \citenamefont
  {Loss}(2012)}]{Klinovaja:PRB12}%
  \BibitemOpen
  \bibfield  {author} {\bibinfo {author} {\bibfnamefont {J.}~\bibnamefont
  {Klinovaja}}\ and\ \bibinfo {author} {\bibfnamefont {D.}~\bibnamefont
  {Loss}},\ }\href {\doibase 10.1103/PhysRevB.86.085408} {\bibfield  {journal}
  {\bibinfo  {journal} {Phys. Rev. B}\ }\textbf {\bibinfo {volume} {86}},\
  \bibinfo {pages} {085408} (\bibinfo {year} {2012})}\BibitemShut {NoStop}%
\bibitem [{\citenamefont {Prada}\ and\ \citenamefont
  {Sols}(2004)}]{Prada:EPJB04}%
  \BibitemOpen
  \bibfield  {author} {\bibinfo {author} {\bibfnamefont {E.}~\bibnamefont
  {Prada}}\ and\ \bibinfo {author} {\bibfnamefont {F.}~\bibnamefont {Sols}},\
  }\href@noop {} {\bibfield  {journal} {\bibinfo  {journal} {Eur. Phys. J. B}\
  }\textbf {\bibinfo {volume} {40}},\ \bibinfo {pages} {379} (\bibinfo {year}
  {2004})}\BibitemShut {NoStop}%
\bibitem [{\citenamefont {Sticlet}\ \emph {et~al.}(2012)\citenamefont
  {Sticlet}, \citenamefont {Bena},\ and\ \citenamefont
  {Simon}}]{Sticlet:PRL12}%
  \BibitemOpen
  \bibfield  {author} {\bibinfo {author} {\bibfnamefont {D.}~\bibnamefont
  {Sticlet}}, \bibinfo {author} {\bibfnamefont {C.}~\bibnamefont {Bena}}, \
  and\ \bibinfo {author} {\bibfnamefont {P.}~\bibnamefont {Simon}},\ }\href
  {\doibase 10.1103/PhysRevLett.108.096802} {\bibfield  {journal} {\bibinfo
  {journal} {Phys. Rev. Lett.}\ }\textbf {\bibinfo {volume} {108}},\ \bibinfo
  {pages} {096802} (\bibinfo {year} {2012})}\BibitemShut {NoStop}%
\bibitem [{Note2()}]{Note2}%
  \BibitemOpen
  \bibinfo {note} {In the continuum limit the Majorana wave function vanishes
  at the ends of the nanowire, $u_\sigma ^{(L,R)}(0)=0$. In that case, the
  hopping amplitudes $t_{L\sigma ,R\sigma }$ are proportional to the spatial
  derivatives $\partial _x u_\sigma ^{(L,R)}(x=0)$ times a lengthscale
  associated to the contact \cite {Prada:EPJB04}. In our minimal contact model
  within a discretized nanowire tight-binding description, such lengthscale is
  the lattice constant $a_0$. With this choice, the hopping amplitudes becomes
  proportional to the Majorana amplitudes at the leftmost site of the
  discretised nanowire, which are no longer zero, but $a_0
  {u'_0}^{(L,R)}(0)$.}\BibitemShut {Stop}%
\bibitem [{Den()}]{Deng:inprep}%
  \BibitemOpen
  \href@noop {} {}\bibinfo {note} {M. Deng et al., in preparation}\BibitemShut
  {NoStop}%
\bibitem [{\citenamefont {Ben-Shach}\ \emph {et~al.}(2015)\citenamefont
  {Ben-Shach}, \citenamefont {Haim}, \citenamefont {Appelbaum}, \citenamefont
  {Oreg}, \citenamefont {Yacoby},\ and\ \citenamefont
  {Halperin}}]{Ben-Shach:PRB15}%
  \BibitemOpen
  \bibfield  {author} {\bibinfo {author} {\bibfnamefont {G.}~\bibnamefont
  {Ben-Shach}}, \bibinfo {author} {\bibfnamefont {A.}~\bibnamefont {Haim}},
  \bibinfo {author} {\bibfnamefont {I.}~\bibnamefont {Appelbaum}}, \bibinfo
  {author} {\bibfnamefont {Y.}~\bibnamefont {Oreg}}, \bibinfo {author}
  {\bibfnamefont {A.}~\bibnamefont {Yacoby}}, \ and\ \bibinfo {author}
  {\bibfnamefont {B.~I.}\ \bibnamefont {Halperin}},\ }\href {\doibase
  10.1103/PhysRevB.91.045403} {\bibfield  {journal} {\bibinfo  {journal} {Phys.
  Rev. B}\ }\textbf {\bibinfo {volume} {91}},\ \bibinfo {pages} {045403}
  (\bibinfo {year} {2015})}\BibitemShut {NoStop}%
\bibitem [{\citenamefont {Dom{\'\i}nguez}\ \emph {et~al.}(2016)\citenamefont
  {Dom{\'\i}nguez}, \citenamefont {Cayao}, \citenamefont {San-Jose},
  \citenamefont {Aguado}, \citenamefont {Yeyati},\ and\ \citenamefont
  {Prada}}]{Dominguez:16}%
  \BibitemOpen
  \bibfield  {author} {\bibinfo {author} {\bibfnamefont {F.}~\bibnamefont
  {Dom{\'\i}nguez}}, \bibinfo {author} {\bibfnamefont {J.}~\bibnamefont
  {Cayao}}, \bibinfo {author} {\bibfnamefont {P.}~\bibnamefont {San-Jose}},
  \bibinfo {author} {\bibfnamefont {R.}~\bibnamefont {Aguado}}, \bibinfo
  {author} {\bibfnamefont {A.~L.}\ \bibnamefont {Yeyati}}, \ and\ \bibinfo
  {author} {\bibfnamefont {E.}~\bibnamefont {Prada}},\ }\href
  {https://arxiv.org/abs/1609.01546} {\  (\bibinfo {year} {2016})},\ \Eprint
  {http://arxiv.org/abs/arXiv:1609.01546} {arXiv:1609.01546} \BibitemShut
  {NoStop}%
\bibitem [{\citenamefont {Lim}\ \emph {et~al.}(2012)\citenamefont {Lim},
  \citenamefont {Serra}, \citenamefont {L\'opez},\ and\ \citenamefont
  {Aguado}}]{Lim:PRB12}%
  \BibitemOpen
  \bibfield  {author} {\bibinfo {author} {\bibfnamefont {J.~S.}\ \bibnamefont
  {Lim}}, \bibinfo {author} {\bibfnamefont {L.}~\bibnamefont {Serra}}, \bibinfo
  {author} {\bibfnamefont {R.}~\bibnamefont {L\'opez}}, \ and\ \bibinfo
  {author} {\bibfnamefont {R.}~\bibnamefont {Aguado}},\ }\href {\doibase
  10.1103/PhysRevB.86.121103} {\bibfield  {journal} {\bibinfo  {journal} {Phys.
  Rev. B}\ }\textbf {\bibinfo {volume} {86}},\ \bibinfo {pages} {121103}
  (\bibinfo {year} {2012})}\BibitemShut {NoStop}%
\bibitem [{\citenamefont {Das~Sarma}\ \emph {et~al.}(2012)\citenamefont
  {Das~Sarma}, \citenamefont {Sau},\ and\ \citenamefont
  {Stanescu}}]{Das-Sarma:PRB12}%
  \BibitemOpen
  \bibfield  {author} {\bibinfo {author} {\bibfnamefont {S.}~\bibnamefont
  {Das~Sarma}}, \bibinfo {author} {\bibfnamefont {J.~D.}\ \bibnamefont {Sau}},
  \ and\ \bibinfo {author} {\bibfnamefont {T.~D.}\ \bibnamefont {Stanescu}},\
  }\href {\doibase 10.1103/PhysRevB.86.220506} {\bibfield  {journal} {\bibinfo
  {journal} {Phys. Rev. B}\ }\textbf {\bibinfo {volume} {86}},\ \bibinfo
  {pages} {220506} (\bibinfo {year} {2012})}\BibitemShut {NoStop}%
\bibitem [{\citenamefont {Rainis}\ \emph {et~al.}(2013)\citenamefont {Rainis},
  \citenamefont {Trifunovic}, \citenamefont {Klinovaja},\ and\ \citenamefont
  {Loss}}]{Rainis:PRB13}%
  \BibitemOpen
  \bibfield  {author} {\bibinfo {author} {\bibfnamefont {D.}~\bibnamefont
  {Rainis}}, \bibinfo {author} {\bibfnamefont {L.}~\bibnamefont {Trifunovic}},
  \bibinfo {author} {\bibfnamefont {J.}~\bibnamefont {Klinovaja}}, \ and\
  \bibinfo {author} {\bibfnamefont {D.}~\bibnamefont {Loss}},\ }\href {\doibase
  10.1103/PhysRevB.87.024515} {\bibfield  {journal} {\bibinfo  {journal} {Phys.
  Rev. B}\ }\textbf {\bibinfo {volume} {87}},\ \bibinfo {pages} {024515}
  (\bibinfo {year} {2013})}\BibitemShut {NoStop}%
\bibitem [{\citenamefont {Jellinggaard}\ \emph {et~al.}(2016)\citenamefont
  {Jellinggaard}, \citenamefont {Grove-Rasmussen}, \citenamefont {Madsen},\
  and\ \citenamefont {Nyg\aa{}rd}}]{Jellinggaard:PRB16}%
  \BibitemOpen
  \bibfield  {author} {\bibinfo {author} {\bibfnamefont {A.}~\bibnamefont
  {Jellinggaard}}, \bibinfo {author} {\bibfnamefont {K.}~\bibnamefont
  {Grove-Rasmussen}}, \bibinfo {author} {\bibfnamefont {M.~H.}\ \bibnamefont
  {Madsen}}, \ and\ \bibinfo {author} {\bibfnamefont {J.}~\bibnamefont
  {Nyg\aa{}rd}},\ }\href {\doibase 10.1103/PhysRevB.94.064520} {\bibfield
  {journal} {\bibinfo  {journal} {Phys. Rev. B}\ }\textbf {\bibinfo {volume}
  {94}},\ \bibinfo {pages} {064520} (\bibinfo {year} {2016})}\BibitemShut
  {NoStop}%
\bibitem [{\citenamefont {Gharavi}\ \emph {et~al.}(2016)\citenamefont
  {Gharavi}, \citenamefont {Hoving},\ and\ \citenamefont
  {Baugh}}]{Gharavi:PRB16}%
  \BibitemOpen
  \bibfield  {author} {\bibinfo {author} {\bibfnamefont {K.}~\bibnamefont
  {Gharavi}}, \bibinfo {author} {\bibfnamefont {D.}~\bibnamefont {Hoving}}, \
  and\ \bibinfo {author} {\bibfnamefont {J.}~\bibnamefont {Baugh}},\ }\href
  {\doibase 10.1103/PhysRevB.94.155417} {\bibfield  {journal} {\bibinfo
  {journal} {Phys. Rev. B}\ }\textbf {\bibinfo {volume} {94}},\ \bibinfo
  {pages} {155417} (\bibinfo {year} {2016})}\BibitemShut {NoStop}%
\bibitem [{\citenamefont {Hoffman}\ \emph {et~al.}(2016)\citenamefont
  {Hoffman}, \citenamefont {Schrade}, \citenamefont {Klinovaja},\ and\
  \citenamefont {Loss}}]{Hoffman:PRB16}%
  \BibitemOpen
  \bibfield  {author} {\bibinfo {author} {\bibfnamefont {S.}~\bibnamefont
  {Hoffman}}, \bibinfo {author} {\bibfnamefont {C.}~\bibnamefont {Schrade}},
  \bibinfo {author} {\bibfnamefont {J.}~\bibnamefont {Klinovaja}}, \ and\
  \bibinfo {author} {\bibfnamefont {D.}~\bibnamefont {Loss}},\ }\href {\doibase
  10.1103/PhysRevB.94.045316} {\bibfield  {journal} {\bibinfo  {journal} {Phys.
  Rev. B}\ }\textbf {\bibinfo {volume} {94}},\ \bibinfo {pages} {045316}
  (\bibinfo {year} {2016})}\BibitemShut {NoStop}%
\bibitem [{\citenamefont {Clarke}(2017)}]{Clarke:17}%
  \BibitemOpen
  \bibfield  {author} {\bibinfo {author} {\bibfnamefont {D.~J.}\ \bibnamefont
  {Clarke}},\ }\href {https://arxiv.org/abs/1702.01740} {\  (\bibinfo {year}
  {2017})},\ \Eprint {http://arxiv.org/abs/arXiv:1702.01740} {arXiv:1702.01740}
  \BibitemShut {NoStop}%
\end{thebibliography}%

\appendix
\section{Parameter relations between tight-binding and effective models}
\label{sec:micro}

To finish our study, we wish to relate analytically the spin canting of the inner Majorana at $x=0$, $\theta_L$, with the microscopic parameters of the nanowire, that is, $\alpha$, $B$, $\mu$ and $\Delta$. This will allow us to quantify the spin-orbit coupling of the nanowire by analysing dot-Majorana splittings. To find this connection, we calculate the Majorana wave function at the end of a semi-infinite wire, see Eq. \eqref{theta}. Following Lutchyn et al. \cite{Lutchyn:PRL10} 
we solve the Bogoliubov-de Gennes (BdG) equations for the nanowire Hamiltonian, $H_w\Psi_{\pm}(x)=\pm E \Psi_{\pm}(x)$, and look for zero energy solutions. These solutions only exist for $B^2>\Delta^2+\mu^2$ and correspond to zero energy Majorana bound states. The eigenstates $ \Psi_{\pm}(x)$ are four component Nambu spinors $\Psi_{\pm}(x)=(u_{\uparrow}(x),u_{\downarrow}(x),v_{\uparrow}(x),v_{\downarrow}(x))^T$ related by the electron-hole symmetry $\Psi_+(x)=S_{eh}\Psi_-(x)$ where $S_{eh}=\sigma_0\tau_x K$, and $K$ is the conjugation operator, $K\Psi(x)=\Psi^*(x)$. Since the BdG Hamiltonian is real we can construct real Nambu spinors $\Psi_{\pm}(x)$. Due to the relation between electron-hole operators and Majorana operators, we can write these spinors in terms of the Majorana left and right spinors: $\Psi_+(x)=[\Psi_L(x)+i\Psi_R(x)]/\sqrt{2}$ and $\Psi_-(x)=[\Psi_L(x)-i\Psi_R(x)]/\sqrt{2}$, where $\Psi_L(x)$ has real components and $\Psi_R(x)$ pure imaginary ones. Due to the Majorana reality condition, Majorana spinors are eigenstates of $S_{eh}$: $S_{eh}\Psi_{L,R}=\lambda \Psi_{L,R}$ with eigenvalue $\lambda=1$ for the left Majorana and $\lambda=-1$ for the right one. This imposes a constraint $(v^{(L,R)}_{\uparrow,\downarrow})^*=\lambda u^{(L,R)}_{\uparrow,\downarrow}$ between hole-like and electron-like components. The 4x4 BdG matrix is then reduced to a 2x2 one: 
\begin{eqnarray*}
\left(\begin{array}{cc}
-\frac{\hbar^2}{2m}\partial_x^2-\mu+B & -\alpha\partial_x+\lambda\Delta \\
\alpha\partial_x-\lambda\Delta  &  -\frac{\hbar^2}{2m}\partial_x^2-\mu-B
\end{array}\right)\left(\begin{array}{c}
u^{(L,R)}_\uparrow(x)\\u^{(L,R)}_\downarrow(x)
\end{array}\right)=0.
\end{eqnarray*}

Since we are interested in the left MBS located at $x=0$ that decays exponentially for $x>0$, we solve this set of coupled differential equations for $\lambda=1$ using the ansatz $u^{(L)}_{\uparrow,\downarrow}(x)\propto u^{(L)}_{\uparrow,\downarrow}e^{-\kappa x}$, with $\textrm{Re}[\kappa]>0$. The characteristic equation for $\kappa$ is a fourth order polynomial with real coefficients
\begin{equation}\label{polynomial}
\left(\frac{\hbar^2}{2m}\right)^2 \kappa^4+\left(\alpha^2+\mu\frac{\hbar^2}{2m}\right)\kappa^2+2\Delta\alpha \kappa+C_0=0,
\end{equation}
where $C_0\equiv\mu^2+\Delta^2-B^2$. As explained by Lutchyn et al. \cite{Lutchyn:PRL10}, it is only possible to find a Majorana wave function that satisfies the boundary condition $\Psi_L(x)=0$ and normalization if $C_0<0$, i.e., in the topological regime. In this case it is possible to express the four solutions of the polynomial in terms of two real positive constants $a$ and $b$ in the following way: $\kappa_{1,2}=a\pm ib$ and $\kappa_{3,4}=-a\pm\sqrt{a^2-4C_0/(a^2+b^2)}$. With this parametrization, only the first three roots have $\textrm{Re}[\kappa_i]>0$.

We can thus write the Majorana wave function as 
\begin{eqnarray}\label{PsiL(x)}
\Psi_L(x)=\left(\begin{array}{c}
u^{(L)}_{\uparrow}(x)\\u^{(L)}_{\downarrow}(x)
\end{array}\right)=\sum_{i=1}^3 A_i \left(\begin{array}{c}
u^{(L)}_{i\uparrow}\\u^{(L)}_{i\downarrow}
\end{array}\right)e^{-\kappa_ix},
\end{eqnarray}
where
\begin{eqnarray}
\left(\begin{array}{c}
u^{(L)}_{i\uparrow}\\u^{(L)}_{i\downarrow}
\end{array}\right)\propto\left(\begin{array}{c}
\frac{\hbar^2}{2m}\kappa_i^2+\mu+B\\-\left(\alpha\kappa_i+\Delta\right)
\end{array}\right).
\end{eqnarray}

Note that Eq. (\ref{PsiL(x)}) exhibits only a two-exponential decay for increasing x, since both $\textrm{Re}[\kappa_1]=\textrm{Re}[\kappa_2]=a$. The three coefficients $A_i$ can be worked out by the two boundary conditions, one for each spinor component, and imposing normalization.

In principle, we can find $\theta_L$ from Eq. (\ref{theta}) and (\ref{PsiL(x)}) by relating the spin-up and -down amplitudes: $\tan(\theta_L/2)=\lim_{x\rightarrow 0}-u_\uparrow^{(L)}(x)/u_\downarrow^{(L)}(x)$. However, since strictly at $x=0$ the bound state wave function is zero (by construction), we have to go to the first derivative to find this relation:
\begin{equation}
\tan\frac{\theta_L}{2}=\left.-\frac{\partial_x u_\uparrow^{(L)}(x)}{\partial_x u_\downarrow^{(L)}(x)}\right|_{x=0}=-\frac{\sum_{i=1}^3\kappa_iA_iu^{(L)}_{i\uparrow}}{\sum_{i=1}^3\kappa_iA_iu^{(L)}_{i\downarrow}}.
\end{equation}
This definition is consistent with Eq. \eqref{theta}. After some algebra we find
\begin{equation}\label{thetaL}
\tan\frac{\theta_L}{2}=-\frac{\frac{\hbar^2}{2m}\kappa_4^2+\mu-B}{\alpha \kappa_4+\Delta},
\end{equation}
where $\kappa_4$ is the real negative root of Eq. (\ref{polynomial}), i.e., precisely the one that doesn't appear in the Majorana wave function Eq. (\ref{PsiL(x)}). Fig. \ref{fig:last}c shows the evolution of the above expression for increasing $B$ and $\alpha$ at $\mu=0$ (solid lines), and a comparison to numerical results using the full tight-binding model (circles).

It is possible to find a manageable analytical solution for $\kappa_4$ in the limit of weak spin-orbit coupling:
\begin{equation}
\kappa_4\sim -\sqrt{\frac{2m}{\hbar^2}}\sqrt{\mu_c-\mu}-\frac{m\alpha \Delta}{\hbar^2\mu_c}+\mathcal{O}(\alpha^2),
\end{equation}
where $\mu_c=\sqrt{B^2-\Delta^2}$. In this limit:
\begin{equation}\label{weakso}
\tan\frac{\theta_L}{2}\sim \frac{B-\mu_c}{\Delta}+\sqrt{\frac{2m}{\hbar^2}}\alpha\frac{\sqrt{\mu_c-\mu}}{\Delta^2}\left(B-\frac{B^2}{\mu_c}\right)+\mathcal{O}(\alpha^2)
\end{equation}
This expression yields a good description of the canting angle at realistic values of $\alpha$, see Fig. \ref{fig:last}c, dashed lines. Note also that if $B$ is much bigger than $\Delta$, the spin canting vanishes as $\theta_L\approx\Delta/B$.

For completeness, we also derive an expansion for strong spin-orbit coupling:
\begin{equation}
\kappa_4\sim -\frac{\Delta+\Delta_c}{\alpha}+\frac{\hbar^2}{2m}\frac{(\Delta+\Delta_c)^2}{\alpha^3\Delta_c}+\mathcal{O}(\alpha^{-5}),
\end{equation}
and
\begin{eqnarray}
\tan\frac{\theta_L}{2} &\sim& \frac{\Delta_c}{B+\mu}+\frac{\hbar^2}{2m\alpha^2}\frac{B(B^2+\Delta^2-\mu^2+2\Delta\Delta_c)}{\Delta_c(B+\mu)} \nonumber\\
&&+\mathcal{O}(\alpha^{-4}),
\end{eqnarray}
where $\Delta_c=\sqrt{B^2-\mu^2}$. We note that, while this expansion does indeed recover the  asymptotic behaviour of Eq. \eqref{thetaL} for large spin-orbit, its regime of validity requires an unphysically large $\alpha$ for real nanowires.

The expressions in this section, together with the measurement scheme for $\theta_L$ encoded in Eq. \eqref{thetaLratio}, provide a powerful method to extract important quantities of the nanowire system, such as spin-orbit coupling $\alpha$.

\end{document}